\documentclass[]{emulateapj}
\usepackage{graphicx, amsmath, amsthm, amssymb}

\shorttitle{ Kozai cycles with tidal friction }
\shortauthors{Fabrycky and Tremaine}

\begin{document}
\slugcomment{Submitted to ApJ}

\title{Shrinking binary and planetary orbits by Kozai cycles with tidal friction}

\author{Daniel Fabrycky\altaffilmark{1} and Scott Tremaine\altaffilmark{1,2}}
\altaffiltext{1}{ Princeton University Observatory, Princeton, NJ 08544 }
\altaffiltext{2}{ Institute for Advanced Study, Princeton, NJ 08540 }
\email{dfab@astro.princeton.edu} 

\begin{abstract}
At least two arguments suggest that the orbits of a large fraction of binary stars and extrasolar planets shrank by 1--2 orders of magnitude after formation: (i) the physical radius of a star shrinks by a large factor from birth to the main sequence, yet many main-sequence stars have companions orbiting only a few stellar radii away, and (ii) in current theories of planet formation, the region within $\sim 0.1$~AU of a protostar is too hot and rarefied for a Jupiter-mass planet to form, yet many ``hot Jupiters'' are observed at such distances.  We investigate orbital shrinkage by the combined effects of secular perturbations from a distant companion star (Kozai oscillations) and tidal friction. We integrate the relevant equations of motion to predict the distribution of orbital elements produced by this process.  Binary stars with orbital periods of 0.1 to 10~days, with a median of $\sim 2$~d, are produced from binaries with much longer periods (10~d to $\sim 10^5$~d), consistent with observations indicating that most or all short-period binaries have distant companions (tertiaries).  We also make two new testable predictions: (1) For periods between $3$ and $10$~d, the distribution of the mutual inclination between the inner binary and the tertiary orbit should peak strongly near $40^\circ$ and $140^\circ$. (2) Extrasolar planets whose host stars have a distant binary companion may also undergo this process, in which case the orbit of the resulting hot Jupiter will typically be misaligned with the equator of its host star. 
\end{abstract}


\section{Introduction}

\subsection{Close binaries with tertiaries}

Close binary star systems (separation comparable to the stellar radii) are often accompanied by a third star.  Such triple-star systems may be stable for long times if the system is hierarchical, that is, if the system consists of an ``inner'' binary (masses $m_1$ and $m_2$) in a nearly Keplerian orbit with semi-major axis $a_{in}$, and an ``outer'' binary in which $m_3$ orbits the center of mass of the inner binary, with semi-major axis $a_{out}\gg a_{in}$.  If the formation of the inner binary is independent of the formation of the outer binary, then the probability that a close binary has a distant companion should be the same as the probability that an individual star has a companion in such an orbit.  It is well known that the latter probability is substantial, about $2/3$ for nearby systems whose brightest component is a G-dwarf \citep{1991DM}. Recent studies suggest that a similarly large fraction of contact binaries have a third component \citep{2006PR, 2006DKR}.

The data show, however, that characteristics of the inner and outer binary in hierarchical triples are {\it not} independent: the probability of having a third component turns out to be a function of the period of the inner binary.  \cite{2006T} found that 96\% of a sample of spectroscopic binaries with periods less than $3$~d had a tertiary component, compared to only 34\% of binaries with periods greater than $12$~d.  This sample was not carefully selected to control biases, however such biases are unlikely to change the basic result, since the authors observe that {\em all} of the five binaries with period less than $10$~d in the volume-limited sample of \cite{1991DM} have at least one additional companion.

A closely related observation is that the period distribution of inner binaries in triple systems is quite different from the period distribution of isolated binaries.  \cite{2002TS} found a significant peak at about $3$~d in the logarithmic period distribution of inner binaries of triple systems, a peak that is not present in isolated binaries.  Because of this feature, the period distributions for binaries with and without a third component are different with a significance of 0.999 \citep{2006T}.  The period distribution from time domain surveys of eclipsing binaries also peaks at a few days \citep{2005D, 2006PSPP, 2007DKB}, but it remains unclear whether this may be attributed to a selection effect: binaries with large orbital periods have a lower probability of eclipsing and fewer eclipses per unit time which diminishes the signal \citep{2005GSM}.

These observational results are surprising.  In particular, the median semi-major axis of the outer binary in systems with inner binary period $<7$~d ($a_{in} \lesssim 0.07$~AU) is $a_{out} \sim 70$~AU \citep{2006T}, similar to the median for all binaries in the \cite{1991DM} sample.  Why are the processes of star formation correlated over a range of three orders of magnitude in scale?  In this paper we explore the possibility that in some circumstances the distant companion enhances tidal interactions in the inner binary, causing its period to shrink to the currently observed value.

\subsection{Kozai cycles}

The study of the long-term behavior of three point masses interacting only through gravity has engaged physicists and mathematicians since the time of Newton.  In most cases, long-term stability requires that the system is hierarchical ($a_{out} \gg a_{in}$).  An additional requirement for a stable hierarchical triple system is that the eccentricity $e_{out}$ of the outer binary cannot be too large, so that $m_3$ cannot make close approaches to $m_1$ or $m_2$. An equivalent statement is that the gravitational perturbations from $m_3$ on the inner binary must always be weak. However, even weak perturbations from the outer body can have important long-term effects on the inner binary. The simplest of these is precession of the orbital plane, which occurs if the orbital planes of the inner and outer binary are not aligned. If the inner and outer binary orbits are circular, this precession is analogous to the precession of two rigid rings with the same mass, radius, and angular momentum as the binary orbits: both the mutual inclination and the scalar angular momenta of the rings remain fixed, while the two angular-momentum vectors precess around the direction defined by the total angular-momentum vector of the triple system.

An unexpected aspect of this behavior was discovered by \cite{1962K}. Suppose the inner binary orbit is initially circular, with the initial mutual inclination between inner and outer binaries equal to $i_{initial}$.  Kozai found that there is a critical angle $i_c$ such that if $i_{initial}$ is between $i_c$ and $180^\circ-i_c$, then the orbit of the inner binary cannot remain circular as it precesses: both the eccentricity of the inner binary $e_{in}$ and the mutual inclination $i$ execute periodic oscillations known as Kozai cycles.  The amplitude of the eccentricity and inclination oscillations is independent of the strength of the perturbation from the outer body, which depends on $m_3$, $a_{out}$, and $e_{out}$, but the oscillation amplitude does depend on $i_{initial}$: for initial circular orbits with $i_{initial} = i_c$ or $180^\circ-i_c$, the maximum eccentricity is $0$, but if $i_{initial}=90^\circ$ the maximum eccentricity is unity; i.e., the two inner bodies collide.

Kozai cycles can be investigated analytically by averaging over the orbital phases of the inner and outer binaries \citep{1962K,2000FKR}; this averaging, usually called the secular approximation, is justified because the precession time is generally much longer than the orbital time of either binary. In the averaged problem the semi-major axes of the inner and outer binary are both conserved. The analysis is simplest in the limiting case when $a_{out}\gg a_{in}$ (so that the perturbing potential of the outer body can be written in the quadrupole approximation) and the angular momentum of the outer binary is much greater than that of the inner binary (so that the orientation of the outer binary is a constant of the motion). With these approximations, the following results hold.  (i) The averaged quadrupole potential from the outer binary is axisymmetric relative to its orbital plane.  (ii) The averaged problem can be described by a Hamiltonian with one degree of freedom.  (iii) The eccentricity of the outer binary is constant.  (iv) The critical inclination is $i_c=\cos^{-1}\sqrt{3/5}=39.2^\circ$.  (v) If the inner orbit is initially circular the maximum eccentricity achieved in a Kozai cycle is $e_{in, max}=[1-(5/3)\cos^2 i_{initial}]^{1/2}$.  (vi) Depending on the initial conditions, the argument of pericenter $\omega_{in}$ (the angle measured in the orbital plane between the pericenter of the inner binary and the orbital plane of the outer binary) can either librate (oscillate around $90^\circ$ or $270^\circ$) or circulate.  The system may remain at a fixed point with $\omega_{in} = 90^\circ$ or $270^\circ$ and $e_{in} = [1-(5/3)\cos^2 i_{fix}]^{1/2}$ if $i_c < i_{fix} < 180^\circ-i_c$.  (vii) The only property of the Kozai oscillation that depends on the masses of the three bodies, their semi-major axes, or the eccentricity of the outer binary is the period of the oscillation, which is of order the timescale \citep{1998KEM}:
\begin{equation}
\tau = \frac{2 P_{out}^2}{ 3 \pi P_{in}} \frac{m_1 + m_2 + m_3}{m_3} (1-e_{out}^2)^{3/2} ;   \label{tau}
\end{equation}
small-amplitude libration about the fixed point takes place with a period
\begin{equation}
P_{lib} = \tau \frac{2 \pi} { \sqrt{30} ( 1 - (5/3)\cos^2 i_{fix} )^{1/2} \sin i_{fix} }.
\end{equation}
(viii) Octupole and higher-order terms in the perturbing potential introduce a narrow chaotic zone around the separatrix between circulating and librating solutions as determined by the quadrupole approximation \citep{1997HTT}.

Consider a sequence of triple systems in which the semi-major axis $a_{out}$ of the outer binary becomes larger and larger, while its mass, inclination, and eccentricity remain the same. The maximum eccentricity of the inner binary in the Kozai cycle will remain fixed, but the period of the Kozai cycle will grow as $a_{out}^3$. This behavior will continue so long as the perturbation from the outer body is the dominant cause of apsidal precession in the inner binary orbit.  Thus, weak perturbations from distant third bodies can induce large eccentricity and inclination oscillations.  However, small additional sources of apsidal precession in the inner binary---general relativity, tides, the quadrupole moments of the two members of the inner binary, planetary companions, etc.---can completely suppress Kozai oscillations caused by a distant third body if they dominate the apsidal precession. 

\subsection{Tides} \label{intro_tides}

The dissipative forces due to tides on the stars of the inner binary are only significant if the two stars are separated by less than a few stellar radii. Thus the tidal friction on an inner binary with a semi-major axis of (say) $0.25$~AU, corresponding to an orbital period of $32$~d if $m_1=m_2=1M_\odot$, is normally negligible. However, if the amplitude of the eccentricity oscillation during a Kozai cycle is sufficiently large, the pericenter distance of the inner binary may become sufficiently small at some phase of the cycle that tidal friction drains energy from the orbit, reducing the semi-major axis and thereby enhancing the friction, until the inner binary settles into a circular orbit with a semi-major axis of only a few stellar radii.  Following \cite{2006T}, we refer to this process as Kozai cycles with tidal friction (KCTF).

\subsection{Previous work}

\cite{1968H} first suggested that KCTF might be an important evolutionary mechanism for triple stars.  \cite{1979MS} showed how the long-period perturbations of a third body could reduce the inner binary's separation on the tidal dissipation timescale.  \cite{1998KEM} used KCTF to constrain the strength of tidal dissipation in stars and focused on the possibility that the inner binary of the Algol system might have shrunk significantly by this mechanism. \cite{2001EK} presented the differential equations that we use here to model KCTF, which parametrize the extra forces (tidal friction, quadrupole from a distant companion, etc.) acting on a binary orbit.

Kozai cycles have also been studied in the planetary context. \cite{1997MKR}, \cite{1997I}, and \cite{1997HTT} all suggested that the large eccentricities observed in many extrasolar planet orbits could be explained by Kozai cycles if the host star were a member of a binary system; however, this hypothesis leads to two predictions that are not verified \citep{2004TZ}: (i) high-eccentricity planets should mostly be found in binary systems; (ii) multi-planet systems should have low eccentricities, since their mutual apsidal precession suppresses the Kozai cycle. Also, \cite{2005TR} argued that even if every extrasolar planet host has an undetected companion, Kozai cycles alone cannot explain the distribution of observed eccentricities.  (They found that too many planets remained on nearly circular orbits, but that result is sensitive to the initial eccentricity assumed since the time spent near the unstable fixed points at $e_{in} = 0$, $\omega_{in} = \pm45^\circ, \pm 135^\circ$ dominates the period of the Kozai cycle, and this time depends strongly on the initial eccentricity; see \citealt{1997I}.) \cite{1998KEM} speculated that KCTF might explain the presence of massive planets on orbits close to their parent stars (``hot Jupiters''), a possibility that we re-examine in this paper. \cite{2003WM} elaborated the speculation of \cite{1998KEM} for the orbit of the planet HD 80606b, which is unusual because of its very large eccentricity ($e=0.93$) and small pericenter distance ($a (1-e)=0.033$~AU).

\cite{2002BLS} have examined a process similar to KCTF for triple black-hole systems that might be found in the centers of galaxies; here the dissipative force is gravitational radiation rather than tidal friction but much of the formalism is the same.  A major difference is that tidal friction vanishes in a binary with a circular orbit and synchronously rotating stars, while gravitational radiation does not.  Therefore the end-state of a black-hole triple subject to the analog of KCTF is a merger, leaving a binary black hole (which may become unbound by the gravitational radiation recoil of the merger).  In the present application, however, gravitational radiation is negligible.  Therefore the stellar and planetary cases offer opportunities to compare the orbits of the observed systems to the distributions predicted by the theory, which this paper quantifies.

\subsection{The plan of this paper}

In this paper we shall model KCTF using (i) the secular approximation for orbital evolution; (ii) the quadrupole approximation for the tidal field from the outer body; (iii) a simple model for apsidal precession that includes the dominant general relativistic precession term and the quadrupole distortion of the stars of the inner binary due to tides and rotation; (iv) a simple model for orbital decay due to tidal friction; (v) the assumption that the outer binary contains most of the angular momentum in the system. Our main goal is to characterize the statistical properties of the binary systems that result from KCTF.  In \S\ref{eom} we describe the equations of motion we use to evolve hierarchical systems under KCTF.  In \S\ref{analsect} we derive a Hamiltonian formalism and the conserved quantities of the system and thereby offer analytical insight into the behavior of Kozai cycles.   We evaluate the effect of tidal friction on a population of isolated binary stars in \S\ref{sec:isobin}.  We numerically integrate a large number of triple systems in \S\ref{mc-calc}, compare the final conditions to an empirical period distribution, and give a prediction for the mutual inclination distribution.  In \S\ref{finali} we verify that the mutual inclination distribution, once established by KCTF, will persist for the lifetime of the system, at least for isolated triple systems.  We apply our results to hot Jupiters orbiting one component of wide binary systems in \S\ref{hotjup}.  We discuss applications and possible extensions of our analysis in \S\ref{sec:discuss} and restate our main conclusions in \S\ref{sec:conclude}.

\section{Equations of motion} \label{eom}
The differential equations that govern the inner binary's orbital parameters and the two stars' spin parameters were presented by \cite{2001EK} and are recalled here.  The equations take into account stellar distortion due to tides and rotation, tidal dissipation based on the theory of \cite{1998EKH}, relativistic precession, and the secular perturbations of a third body (averaged over the inner and outer Keplerian orbits).  The orientation of the inner orbit is specified by its Laplace-Runge-Lenz vector, ${\bf e}_{in}$, whose magnitude is the inner eccentricity $e_{in}$ and whose direction is towards the pericenter of the orbit of $m_2$ about $m_1$.  Also needed is the specific angular momentum vector ${\bf h}_{in}$ of the inner orbit, whose magnitude is $[G (m_1+m_2) a_{in} (1-e_{in}^2)]^{1/2}$, where $G$ is the gravitational constant.  The vector ${\bf \hat{q}}_{in} = {\bf\hat{h}}_{in} \times {\bf \hat{e}}_{in}$ completes the right-hand triad of unit vectors $( {\bf \hat{e}}_{in},  {\bf \hat{q}}_{in},  {\bf \hat{h}}_{in} )$.  Each of the stars of the inner binary also has a spin vector ${\bf \Omega}_j$, $j=1, 2$.

The evolution equations are:

\begin{multline}
\frac{1}{e_{in}} \frac{d {\bf e}_{in}}{dt} = (Z_1 + Z_2 + Z_{GR}) {\bf \hat{q}}_{in} - (Y_1 + Y_2) {\bf \hat{h}}_{in} -(V_1 + V_2) {\bf \hat{e}}_{in} \\
  - (1-e_{in}^2) [5 S_{eq} {\bf \hat{e}}_{in} -(4 S_{ee} - S_{qq}) {\bf \hat{q}}_{in} + S_{qh} {\bf \hat{h}}_{in} ],  \label{edoteq} 
\end{multline}
\begin{multline}
\frac{1}{h_{in}} \frac{d {\bf h}_{in}}{dt} = (Y_1+Y_2) {\bf \hat{e}}_{in} - (X_1 + X_2) {\bf \hat{q}}_{in} - (W_1 + W_2) {\bf \hat{h}}_{in} \\
+ (1-e_{in}^2) S_{qh} {\bf \hat{e}}_{in} - (4e_{in}^2+1) S_{eh} {\bf \hat{q}}_{in} + 5 e_{in}^2 S_{eq} {\bf \hat{h}}_{in},  
\end{multline}
\begin{eqnarray}
I_1 \frac{d {\bf \Omega}_1 }{dt} &=& \mu h_{in} (-Y_1{\bf \hat{e}}_{in} + X_1 {\bf \hat{q}}_{in} + W_1 {\bf \hat{h}}_{in}), \\
I_2 \frac{d {\bf \Omega}_2 }{dt} &=& \mu h_{in} (-Y_2{\bf \hat{e}}_{in} + X_2 {\bf \hat{q}}_{in} + W_2 {\bf \hat{h}}_{in}).\label{om2doteq} 
\end{eqnarray}

Subscripts $1$ and $2$ refer to the masses $m_1$ and $m_2$.  The gravitational influence of the third body is described by the tensor ${\bf S}$.  We specify the orientation of the orbit of the third body using the triad $({\bf \hat{e}}_{out},  {\bf \hat{q}}_{out},  {\bf \hat{h}}_{out} )$, defined by analogy with the inner binary's coordinate system, treating the inner binary as a point mass $m_1+m_2$ at its center of mass.  We have:

\begin{eqnarray}
S_{mn} = C (\delta_{mn} - 3 \hat{h}_{out,m} \hat{h}_{out,n} ) \label{stensor}\\
C = \frac{ 2 \pi }{ 3 \tau } (1-e_{in}^2)^{-1/2}
\end{eqnarray}

Here the directions $m, n \in ( \hat{e}_{in},  \hat{q}_{in}, \hat{h}_{in} )$ are along the basis vectors of the inner orbit, and $\hat{h}_{out,m}$ and $\hat{h}_{out,n}$ are direction cosines of the outer orbit's angular momentum along the inner orbit's coordinate directions.  $P_{in}$ and $P_{out}$ are the periods of the inner and outer orbits, respectively.  $V$ and $W$ are dissipation rates for $e$ and $h$, respectively.  The vector $(X,Y,Z)$ in the $( {\bf \hat{e}}_{in},  {\bf \hat{q}}_{in},  {\bf \hat{h}}_{in} )$ frame is the angular velocity of that frame relative to the inertial frame.  $Z_{GR}$ is the first post-Newtonian effect of relativity which causes the pericenter to precess.  The stars of the inner binary have moment of inertia $I_1$ and $I_2$, and $\mu = m_1 m_2 / (m_1 + m_2)$ is the reduced mass of the inner binary.  The functional forms of the $V$, $W$, $X$, $Y$, and $Z$ terms were modeled in \cite{2001EK} and for convenience are repeated in the appendix.

\begin{figure*}
\centering
\includegraphics[scale=0.7]{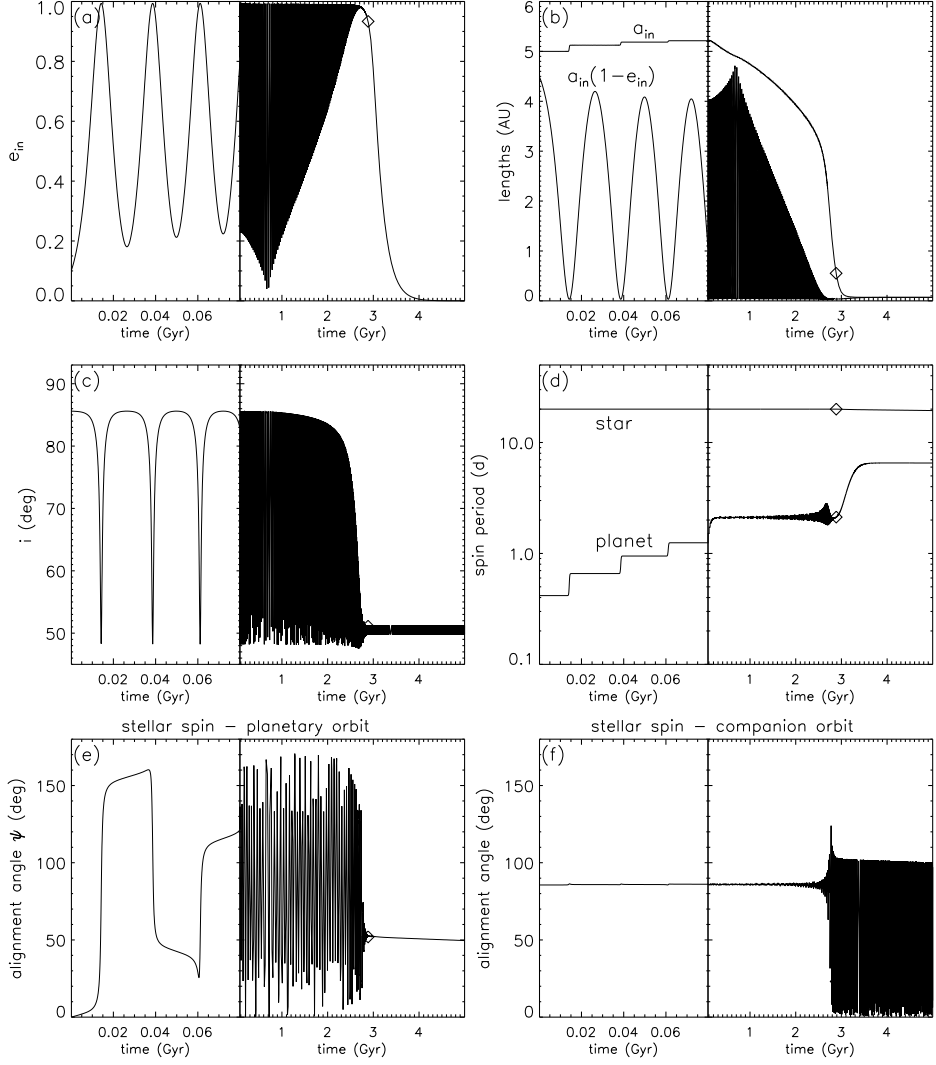}
\caption{The evolution of a planet initially in an orbit with $a_{in}=5$~AU, $i=85.6^{\circ}$, $e_{in}=0.1$, $\omega_{in}=45^{\circ}$, as a hypothetical progenitor to HD 80606b.  The spins of both the planet and its host star were initialized with zero obliquity.  The stellar companion was assumed to have $m_3 = 1.1 m_{\sun}$, $a_{out} = 1000$~AU, and $e_{out} = 0.5$.  The values of structural parameters were the same as those used by \cite{2003WM}, and the viscous times were $t_{V,star} = 50$~yr and $t_{V,planet}=0.001$~yr (see appendix).  Energy dissipation is dominated by the planet.  The diamonds mark the current position of HD 80606 along this possible evolution. \vspace{0.3 in} }
\label{kctf_wm}
\end{figure*}

To check the code we use to integrate these equations, we have replicated the results of \cite{2003WM}, who studied KCTF for the planet HD80606b, whose host star has a binary companion with $a_{out} \sim 10^3$~AU.  We recalculate their Figure 1, using their specified parameters and initial conditions (Fig.~\ref{kctf_wm}).  We started both the planet's and the star's obliquity (the angle between $\hat{\bf h}_{in}$ and ${\bf \Omega}_1$ or ${\bf \Omega}_2$) at $0^\circ$.  Prominent eccentricity oscillations are seen in panel (a).  The energy in the planet's spin was transferred to the orbit, increasing the semi-major axis for the first $0.1$~Gyr (panel b).  A rotational equilibrium was reached which matched the spin angular velocity to that of the instantaneous orbit angular velocity when tides were the strongest, i.e., for pericenter passage during the high-eccentricity part of Kozai cycles.  As dissipation shrunk the semi-major axis, pericenter precession became gradually dominated by relativity rather than by the third body.  Thus, although the orbit initially had a librating pericenter, it started circulating as the eccentricity passed close to zero at $0.7$~Gyr (panel a).  During the Kozai cycles, the orbital angular momentum of the planet is generally quite misaligned from the spin angular momentum of the star (panel e); i.e., the stellar obliquity $\psi$ is substantial.  After $a_{in}$ equilibrates, a large amount of misalignment will generally remain for the main-sequence lifetime of the star, although slight movement towards alignment occurs because of dissipative tides being raised on the star by the planet.  The tidal field of the planet applies a torque to the rotational bulge of its host star, which causes the spin axis to precess; compared to the fixed reference frame of the third body's orbit, the host star's spin axis oscillates with large amplitude (panel f).  To conserve angular momentum, the planet's orbit likewise precesses, but on a considerably smaller scale---it is the few degree wiggle in the inclination after migration (after $2.8$~Gyr in panel c).  We will consider these alignment issues in \S\ref{finali} and \S\ref{RMeffect}.
 
\section{Analytic understanding of KCTF} \label{analsect}

In this section we seek an approximate analytical theory of KCTF.  To lowest order in $a_{in}/a_{out}$, the tertiary component presents a quadrupole tide to the inner binary.  As $a_{in}$ shrinks by KCTF, this approximation will become better and better.  The instantaneous quadrupole-order Hamiltonian is \citep{2000FKR}:
\begin{eqnarray}
\mathcal F &=& -\frac{G m_1 m_2}{2 a_{in}} - \frac{G (m_1 + m_2) m_3}{2 a_{out}} + \mathcal F_q\\
\mathcal F_q &=& -\frac{G m_1 m_2 m_3}{ m_1 + m_2 } \frac{r_{in}^2}{r_{out}^3} P_2(\cos \Phi),
\end{eqnarray}
where ${\bf r}_{in}$ is the vector from $m_1$ to $m_2$, ${\bf r}_{out}$  is the vector from the inner binary center of mass to $m_3$, $\Phi$ is the angle between ${\bf r}_{in}$ and ${\bf r}_{out}$, and $P_2(x) = \frac32 x^2 - \frac12$.

This quadrupole-order potential may be integrated over the unperturbed motion in both orbits to remove short period terms, which depend on the two orbits' mean anomalies.  However, as in \cite{1997I}, we shall work in the approximation that the outer orbit contains essentially all the angular momentum and hence is fixed, therefore the mutual inclination $i$ is the same as the inclination relative to the total angular momentum ($i_{in} = i$, $i_{out} = 0$).  The averaged Hamiltonian is \citep{1997I, 2000FKR}:

\begin{equation}
\langle \mathcal F \rangle = -\frac{G m_1 m_2}{2 a_{in}} - \frac{G (m_1 + m_2) m_3}{2 a_{out}} + \langle \mathcal F_q \rangle  \label{eqn:aveF} 
\end{equation}
\begin{multline}
\langle \mathcal F_q \rangle = -\frac{G m_1 m_2 m_3}{m_1+m_2} \frac{a_{in}^2}{8 a_{out}^3 (1-e_{out}^2)^{3/2}} \\
  \times  \Big{(} 2 + 3e_{in}^2 - (3+12e_{in}^2-15e_{in}^2 \cos^2 \omega_{in} ) \sin^2 i \Big{)}, \label{eqn:fqsecular}
\end{multline}
where the orbital elements are referenced to the plane of the outer binary.  The canonical Delaunay variables for the inner binary are the mean anomaly $l_{in}$, argument of pericenter $\omega_{in}$, and longitude of the ascending node $\Omega_{in}$, along with their respective canonical momenta $L_{in} = m_1 m_2 \sqrt{G a_{in} / (m_1 + m_2)}$, $G_{in} = L_{in} \sqrt{1 - e_{in}^2} = \mu h_{in}$, and $H_{in} = G_{in} \cos i$.  The canonical equations of motion for the inner orbit are:

\begin{eqnarray}
\dot l_{in} = \frac{ \partial \langle \mathcal F \rangle } {\partial L_{in}}, &   \qquad   &\dot L_{in} = -\frac{ \partial \langle \mathcal F \rangle } {\partial l_{in}}, \\
\dot \omega_{in} = \frac{ \partial \langle \mathcal F \rangle } {\partial G_{in}}, &   \qquad    & \dot G_{in} = -\frac{ \partial \langle \mathcal F \rangle } {\partial \omega_{in}}, \label{eqn:hamG} \\
\dot \Omega_{in} = \frac{ \partial \langle \mathcal F \rangle } {\partial H_{in}}, &    \qquad   & \dot H_{in} = -\frac{ \partial \langle \mathcal F \rangle } {\partial \Omega_{in}},
\end{eqnarray}
with corresponding equations for the evolution of the coordinates and momenta of the outer variables.

The averaging procedure removes the Hamiltonian's dependence on the mean anomalies $l_{in}$ and $l_{out}$, so $L_{in}$ and $L_{out}$ and thus the semi-major axes of the inner and outer orbits are conserved.  The average Hamiltonian $\langle \mathcal F \rangle$ is conserved, because the Hamiltonian is independent of time; moreover $\langle \mathcal F_q \rangle$ depends only on $\langle \mathcal F \rangle$ and the conserved semi-major axes (eq.~[\ref{eqn:aveF}]), so $\langle \mathcal F_q \rangle$ is also conserved.  Another conserved quantity is $H_{in}$, because the Hamiltonian is independent of its canonical conjugate, $\Omega_{in}$.  A final conserved quantity is the eccentricity of the outer binary, $e_{out}$, because the Hamiltonian is independent of $\omega_{out}$.

We may write dimensionless versions of the conserved quantities as:

\begin{eqnarray}
F' &=& - 2 - 3e_{in}^2 \nonumber \\
     &  &   + (3+12e_{in}^2-15e_{in}^2 \cos^2 \omega_{in} ) \sin^2 i\\
H' &=& (1-e_{in}^2)^{1/2} \cos i. \label{eqn:hp}
\end{eqnarray}

\cite{1998KEM} have given a different conserved quantity which is simply a combination of the above: $(5-3{H'}^2 +F')/3$.

These constants can be evaluated with the initial set of elements $(e_{in,initial}, \omega_{in,initial}, i_{initial})$.  It is then possible to compute analytically the values of $e_{in}$ and $i$ at any value of $\omega_{in}$ accessible to the system.  Let us restrict our discussion to $i_{initial} > 90^\circ$ (for retrograde systems, identical behavior of the inclination, mirrored across $90^\circ$, results).  The system attains a maximum $e_{in}$ and minimum $i$ when $\omega_{in} = 90^\circ$ or $270^\circ$.  Therefore, we have:

\begin{eqnarray}
e_{in,max} &=& \Big( \{[(10+12{H'}^2-F')^2 - 540{H'}^2]^{1/2} \nonumber \\
     & &   + 8-12{H'}^2+F'  \}/18 \Big)^{1/2} \label{emaxeqn}\\
i_{min} &=& \cos^{-1} [ H' (1-e_{in,max}^2)^{-1/2}] .
\end{eqnarray} 
If $F'$ is initialized with $e_{in}=0$, the value of $e_{in,max}$ given by equation~(\ref{emaxeqn}) matches the value quoted in the introduction.

\begin{figure}
\includegraphics*[scale=0.45]{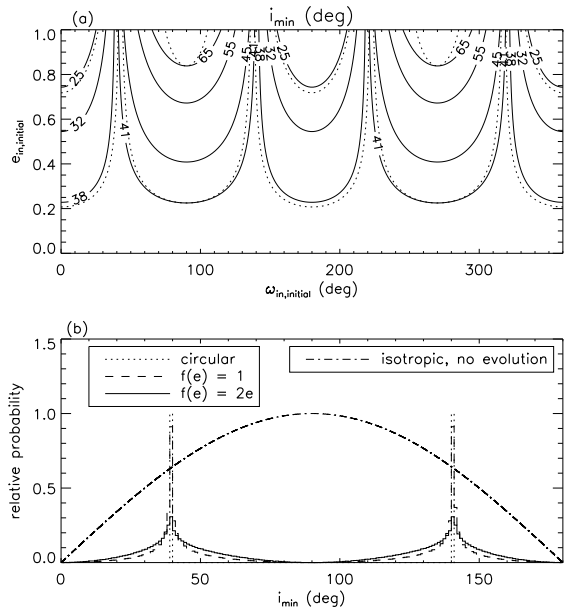}
\caption{(a) The minimum inclination during a Kozai cycle, starting with $i_{initial}=86^\circ$ at various eccentricities $e_{in,initial}$ and arguments of pericenter $\omega_{in,initial}$ for the inner orbit (solid lines).  These results depend very weakly on the initial inclination; for $i_{initial}=70^\circ$, the contours for $i_{min} = 25^\circ$, $38^\circ$, $41^\circ$, and $65^\circ$ (dotted lines) are only slightly different.  (b) The inclination minimum distribution (probability density function), assuming $\omega_{in,initial}$ is uniformly distributed in angle.  These distributions result from three different assumptions for the initial eccentricity distribution: circular orbits, uniform distribution in eccentricity $e_{in,initial}$, and uniform distribution in $e_{in,initial}^2$ (constant phase-space density on the energy surface).  They were computed with $i_{initial} = 86^{\circ}$, but all initial inclinations near $90^\circ$ (i.e., all systems that evolve substantially by KCTF) produce similar distributions.  If KCTF seals in this inclination minimum (see text), the distribution of triples will show spikes at the critical angles $i_c \simeq 40^\circ$ and $180^\circ-i_c \simeq 140^\circ$ and very few systems near $90^\circ$.  An isotropic distribution, with no correlation between the directions of inner and outer orbital angular momenta, is plotted for comparison.  See also Figure~\ref{inclination_dist}b for the inclination distribution that results from integrating the full equations of motion for a population of triples. \vspace{0.3 in} }
\label{contoureoi}
\end{figure}

Figure~\ref{contoureoi}a is a contour plot of the minimum mutual inclination attained as a function of initial eccentricity and argument of pericenter.  For an initial eccentricity $e_{in,initial} < 0.2$, the minimum inclination is within $2^\circ$ of the critical inclination, $i_c = 39.2^{\circ}$.  Integrating over a uniform initial distribution of arguments of pericenter and various initial eccentricity distributions gives the probability distributions of minimum inclination shown in Figure~\ref{contoureoi}b.  These results are only weakly dependent on the initial inclination.

We now argue that the minimum inclinations shown in Figure~\ref{contoureoi}a are very nearly equal to the final inclinations produced by KCTF.  If we approximate the effect of tidal dissipation as acting only when the inclination is at its minimum (i.e., eccentricity maximum), then $i_{min}$ is conserved between Kozai cycles (Fig.~\ref{kctf_wm}c) even though the constants $F'$ and $H'$ are not conserved in the presence of tidal friction.  Eventually, all eccentricity is damped (after $4$~Gyr in Fig.~\ref{kctf_wm}a) and the system finishes with an inclination nearly equal to its minimum inclination from the first cycle (Fig.~\ref{kctf_wm}c).  The distributions of minimum inclination shown in Figure~\ref{contoureoi}b will correspond to the final inclinations after KCTF if the above approximations are correct.  Plotted for comparison is the inclination distribution of an isotropic distribution of triples, i.e., triples with inner and outer angular momentum directions uncorrelated.  The distributions clearly distinguish a population of triples whose inner binaries have shrunk by KCTF from a population with its primordial inclinations, if these are uncorrelated.

This calculation will only be a faithful representation of the final inclination distribution if pericenter precession due to other causes is negligible compared to the precession caused by the tertiary component.  Under these conditions, the final distribution of mutual inclination shows strong peaks close to the critical angle for Kozai cycles: $39.2^\circ$ or $140.8^\circ$, as seen in Figure~\ref{contoureoi}b.  For a more realistic calculation considering a population of triples and including the extra precession forces, see \S\ref{mc-calc} and Figure~\ref{inclination_dist}.

\subsection{Kozai cycles in the presence of extra forces} \label{exforce}

Because Kozai cycles are driven by the interplay between the weak tidal forces from the outer binary and the shape of the orbit of the inner binary, they can be easily suppressed by other weak effects leading to pericenter precession in the inner binary.  Mathematically, the rate of change of eccentricity is proportional to $-S_{eq} = 3 C \hat{h}_{out,e} \hat{h}_{out,q} \propto \sin 2 \omega_{in}$ (according to eq.~[\ref{edoteq}] and eq.~[\ref{stensor}]).  If $\omega_{in}$ is changing too fast, $S_{eq}$ will average to zero before $e_{in}$ has time to grow.  We consider four causes of extra pericenter precession---relativity, tides, stellar rotational distortion, and extra bodies.  Here we characterize the first three effects with additional Hamiltonian terms and calculate the associated apsidal motion $\dot{\omega}_{in}$.  For extra Hamiltonian terms $\langle \mathcal{F}_{extra} \rangle$, equation~(\ref{eqn:hamG}) gives:

\begin{equation}
\dot{\omega}_{extra} = \frac{\partial \langle \mathcal{F}_{extra} \rangle }{\partial G_{in}} = - \frac{ (1-e_{in}^2)^{1/2} }{ e_{in} L_{in} } \frac{\partial \langle \mathcal{F}_{extra} \rangle}{\partial e_{in}}.
\end{equation}

First, let us consider relativity.  As in \cite{2001EK}, we only consider pericenter precession, which is the largest effect.  Take an extra Hamiltonian term of 
\begin{equation}
\langle \mathcal{F}_{GR} \rangle (e_{in}) = - \frac{3 G^2 m_1 m_2 (m_1 + m_2)}{ a_{in}^2 c^2 } \frac{1}{(1-e_{in}^2)^{1/2}}, 
\end{equation}
which depends on both $L_{in}$ and $G_{in}$.  The former dependence gives an additional contribution to the mean motion, which is incorrect but does not affect the secular results.  The latter gives
\begin{equation}
\dot{\omega}_{GR} = \frac{3 G^{3/2} (m_1 + m_2)^{3/2}}{a_{in}^{5/2} c^2 (1-e_{in}^2)},    \label{omgr}
\end{equation}
which is the standard expression for the rate of pericenter precession due to relativistic effects.  

Next let us consider how the non-dissipative tidal bulge contributes to the apsidal motion.  At any moment, construct a spherical polar coordinate system centered on $m_2$ with radius $r_2$ and a polar angle $\theta'$ measured from the vector $\hat{r}_{in}$.  The tidal potential of $m_1$ to lowest order in $r_2/r_{in}$ is
\begin{equation}
\phi_{1,t} = - G m_1 P_2(\cos \theta') r_2^{2} r_{in}^{-3}.
\end{equation}
The surface of $m_2$ is an equipotential.  The corresponding distortion produces an external potential of
\begin{equation}
\phi_{2,ep} = - 2 G m_1 P_2(\cos \theta') r_2^{-3} r_{in}^{-3} R_2^5 k_2, \label{eqn:instphi2ep}
\end{equation}
where $R_{j}$ is the radius and $k_{j}$ is the classical apsidal motion constant \citep{1928R} of the $j$th body.  Typical values of $k_{j}$ are $0.014$ for stars and $0.25$ for gas giant planets.  Back at $m_1$ (for $r_2 = r_{in}$, $\theta'=\pi$), the extra force per unit mass is
\begin{equation}
- \frac{\partial \phi_{2,ep} }{ \partial r_2 } \Big{|}_{r_2=r_{in}} = -6 G m_1 r_{in}^{-7} R_2^5 k_2, \label{eqn:tidalforce}
\end{equation}
which can be integrated to find the potential associated with the tidal distortion of $m_2$ by $m_1$:
\begin{equation}
\phi_{1,2} = - G m_1 r_{in}^{-6} R_2^5 k_2.
\end{equation}
Notice this is a factor of 2 smaller than equation~(\ref{eqn:instphi2ep}) evaluated at the location of $m_1$ because the derivative of $\phi_{2,ep}$ is taken with respect to the spatial coordinate $r_2$, but $\phi_{1,2}$ is assembled by integrating equation~(\ref{eqn:tidalforce}) with respect to $r_{in}$ (see, e.g., \citealt{1939S} and the Appendix of \citealt{2001EK}).  By analogy with the theory of image charges in electrostatics, the effective potential is half the physical potential (P. Eggleton, private communication).  From this potential we may form the instantaneous Hamiltonian $\mathcal{F}_{Tide,2} = m_1 \phi_{1,2}$.  After averaging over one orbit of the inner binary and accounting for both stars, we have
\begin{eqnarray}
\langle \mathcal{F}_{Tide}\rangle(e_{in}) &=& - \frac{G}{8 a_{in}^6 } \frac{8 + 24 e_{in}^2 + 3 e_{in}^4}{(1-e_{in}^2)^{9/2} } \nonumber \\
& & \times \Big[ m_2^2 k_1 R_1^5 + m_1^2 k_2 R_2^5 \Big]   .
\end{eqnarray}
After converting to canonical variables, the equations of motion yield:
\begin{eqnarray}
\dot{\omega}_{Tide} &=& \frac{15 (G (m_1 + m_2))^{1/2} }{8 a_{in}^{13/2} } \frac{8 + 12 e_{in}^2 + e_{in}^4}{(1-e_{in}^2)^5} \nonumber \\
 & &  \times \Big[ \frac{m_2}{m_1} k_1 R_1^5 + \frac{m_1}{m_2} k_2 R_2^5 \Big].
\end{eqnarray}
This expression is always positive, so tidal bulges always tend to promote pericenter precession and therefore suppress Kozai oscillations.  

An analogous procedure gives the extra piece of the Hamiltonian arising from the rotational bulges of the stars of the inner binary.  The instantaneous Hamiltonian resulting from the quadrupole field of the rotationally oblate $m_2$ is:
\begin{equation}
 \mathcal{F}_{Rotate,2} = \mbox{\small{$\frac23$}} k_2 \frac{m_1 \Omega_2^2 R_2^5}{r_{in}^3} P_2(\cos\theta),
\end{equation}
where $\theta$ is the angle measured from the spin axis of $m_2$.  Averaging the result, accounting for both stars and putting $\Omega_1$ and $\Omega_2$ in component form, 
\begin{multline}
\langle \mathcal{F}_{Rotate} \rangle(e_{in}, {\bf \Omega}_1, {\bf \Omega}_2) = - \frac{m_1 m_2 }{6 a_{in}^3 (1-e_{in}^2)^{3/2}}  \\
 \times \sum_{j=1,2}  \frac{k_j R_j^5}{m_j} (2 \Omega_{jh}^2 - \Omega_{je}^2 - \Omega_{jq}^2) .
\end{multline}
To use the canonical equations of motion, each $\Omega_{j}$ component must be converted to components in the inertial frame.  After taking a derivative with respect to $G_{in}$ and converting back to the orbit frame, we have:
\begin{multline}
\dot{\omega}_{Rotate} = \frac{(m_1 + m_2)^{1/2} }{2 G^{1/2} a_{in}^{7/2} (1-e_{in}^2)^2}  \\
  \times \sum_{j=1,2} \frac{k_j R_j^5}{m_j}  \Big[ (2 \Omega_{jh}^2 - \Omega_{je}^2 - \Omega_{jq}^2)  \\
+ 2 \Omega_{jh} \cot i (\Omega_{je} \sin \omega_{in} + \Omega_{jq} \cos \omega_{in}) \Big].
\end{multline}

Small additional bodies in the system, such as planets, also affect
the dynamics through their contribution to the apsidal precession
rates of the stars in the inner binary.  In binary stellar systems,
planetary orbits may stay close to one of the stars (S-type orbits,
for ``satellite'') or encompass the whole binary (P-type orbits, for
``planetary''). Stable S-type or P-type orbits must obey certain
stability criteria \citep{1999HW, 2001MA, 2006MY}; crudely, these
require that the semi-major axis $a\ll a_{in}$ for stable S-type
orbits, while $a\gg a_{in}$ for stable P-type orbits.  In triple
stellar systems, an analogous classification scheme can be worked out.
Three different types of orbits may be stable: (1) S-type about any of
the stars, (2) P-type with respect to the inner binary, but S-type
with respect to the outer binary, and (3) P-type with respect to the
outer binary.  In the first case, the time-average of an S-type orbit
in the equatorial plane of its host star will qualitatively act as an
additional contribution to stellar oblateness.  Although large
eccentricity oscillations of the inner binary would tend to destabilize
S-type planets, the extra pericenter precession caused by the planet
may suppress those oscillations: even a tiny planet may thereby be
responsible for its own survival.  In case (2) the outer binary can
induce Kozai oscillations in the planetary orbit, while in case (3)
the planet can induce Kozai oscillations in the outer binary. The
contribution of such additional bodies to Kozai oscillations has been
studied in the context of multi-planet S-type systems in binaries
\citep{1997I, 2003WM, 2006MDC}, but the huge parameter space of the
general secular four-body problem has not yet been systematically explored.

\subsection{Modified eccentricity maximum} \label{mod-ei}

When the additional forces described in \S\ref{exforce} are present, we may compute $e_{max}$ as before, although the conserved Hamiltonian is now 
\begin{equation}
\langle \mathcal{F}_{tot} \rangle = \langle \mathcal{F}_q \rangle  + \langle {\mathcal F}_{extra} \rangle, 
\end{equation}
where $\langle{\cal F}_{extra}\rangle= \langle{\cal F}_{GR}\rangle+\langle{\cal F}_{Tide}\rangle +\langle{\cal F}_{Rotate}\rangle$.  If $\langle \mathcal{F}_{Rotate} \rangle$ contributes significantly, one must take into account that each component of the spins ($\Omega_{1i}$, $\Omega_{2i}$) will be a function of time.

\begin{figure}
\plotone{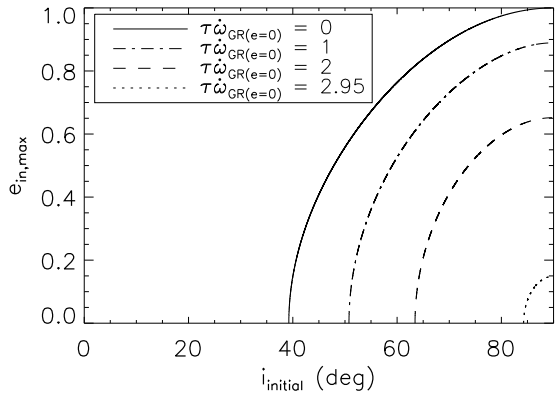}
\caption{The maximum eccentricity attained by systems with initially circular orbits of varying initial inclinations, including relativistic precession (eq.~[\ref{eqn:emaxmod}]).  The curves are parameterized by the relative strength of relativistic precession to that of the tidal field of the third body: $\tau \dot{\omega}_{GR}\big{|}_{e_{in}=0}$, where $\tau$ is defined by equation~(\ref{tau}) and $\dot{\omega}_{GR}$ (eq.~[\ref{omgr}]) is evaluated at $e_{in}=0$. As the timescale of relativistic precession becomes as short as the timescale of precession induced by the companion, the eccentricity cycles are reduced in amplitude and the critical inclination---the largest $i_{initial}$ for which the orbit remains circular---grows (eq.~[\ref{icrit_gr}]).   In contrast to the situation without relativistic precession, a system in which the inner and outer binary orbits are initially perpendicular ($i_{initial} = 90^\circ$) will not reach a radial orbit but attain a more moderate eccentricity (see eq.~[\ref{emax_gr}]).  No Kozai oscillations occur for any initial inclination if $\tau \dot{\omega}_{GR}\big{|}_{e_{in}=0} \geq 3$.\vspace{0.3 in}  }
\label{maxe}
\end{figure}

Let us consider the case in which $\langle {\mathcal F}_{extra} \rangle$ is dominated by $\langle {\mathcal F}_{GR} \rangle$ and assume that the orbit is initially circular to evaluate $\langle {\mathcal F}_{tot} \rangle$.  Eccentricity maximum still occurs at $\omega_{in} = 90^\circ$ or $270^\circ$, which simplifies the expression for $\langle {\mathcal F}_q \rangle$ at eccentricity maximum.  The conservation of $\langle {\mathcal F}_{tot} \rangle$ and $H'$ (eq.~[\ref{eqn:hp}]) gives an implicit equation for $e_{in,max}$:

\begin{multline}
\cos^2 i_{initial} = {\mbox {\small $\frac35$}} (1 - e_{in,max}^2) \\
- {\mbox {\small $\frac25$}} [ (1 - e_{in,max}^2)^{-1/2} + 1]^{-1} \tau \dot{\omega}_{GR} \big{|}_{e_{in}=0} , \label{eqn:emaxmod}
\end{multline}
where the strength of relativity is parametrized by the product of the Kozai timescale, $\tau$ (eq.~[\ref{tau}]) and $\dot{\omega}_{GR}$ evaluated at $e_{in} = 0$ (eq.~[\ref{omgr}]).  We plot this function in Figure~\ref{maxe}.  There is a maximum eccentricity that can be reached for a given amount of GR precession; beginning with mutual inclination of $90^\circ$ and negligible eccentricity:

\begin{equation}
e_{in, max} = \{ 1 - [  ({\mbox {\small $\frac14$}} + {\mbox {\small $\frac23$}} \tau \dot{\omega}_{GR}\big{|}_{e_{in}=0})^{1/2} - 1/2 ]^2 \}^{1/2}. \label{emax_gr}
\end{equation}

The critical inclination for eccentricity oscillations is also increased by relativity:
\begin{equation}
\cos^2 i_{c} = 3/5 - (1/5)\tau \dot{\omega}_{GR}\big{|}_{e_{in}=0}   .\label{icrit_gr}
\end{equation}
In summary, some hierarchical triples, even if they begin with perpendicular orbits, may avoid close encounters because relativistic precession suppresses Kozai cycles.  A similar analysis has been performed by \cite{2002BLS}.

\section{Tidal shrinkage in isolated binaries} \label{sec:isobin}

In this section, we select a sample of isolated binaries via Monte Carlo methods, then we follow how their orbits evolve by tidal dissipation.  This calculation provides a control sample for the numerically integrations of the following section, in which we follow the secular evolution of triple stars by KCTF.  

The initial orbital distributions were taken from the observed distributions of \cite{1991DM}.  For simplicity, $m_1$ was set to $M_\sun$.  Next, $m_2$ was chosen by selecting the mass ratio $q_{in} = m_2/m_1$ from a Gaussian distribution with mean 0.23 and standard deviation 0.42, as found by \cite{1991DM} (negative results were resampled).  The stars were given radii consistent with their masses for stars on the main sequence: $R_1 = R_\sun$ and $R_2 = R_\sun (m_2/M_\sun)^{0.8}$ \citep{1994KW}.  Their structural constants were set to typical values given in \cite{2001EK}: $k_1=k_2=0.014$ (apsidal motion constants), $t_{V1} = t_{V2} = 5$~yr (viscous timescales; see appendix), $I_1 = 0.08 M_1 R_1^2$ and $I_2 = 0.08 M_2 R_2^2$ (moments of inertia).  The initial spin periods were each $10$~d, with spin angular momenta aligned with the binary's orbital angular momentum.  

The orbital period was picked from the log-normal distribution of \cite{1991DM} (with $\langle \log P [\rm{d}] \rangle = 4.8$ and $\sigma_{\log P [\rm{d}]} = 2.8$).  It is likely that a lower cutoff needs to be imposed because the large radii of pre-main-sequence stars likely preclude periods below a few days, and a cutoff of $0.2$~d is seen empirically for main-sequence binaries \citep{2006PSPP}.  We do not impose any cutoff at this stage, but we will explore such cutoffs when presenting the results.

The distribution of initial eccentricities was a function of period, in accord with observed binaries \citep{1991DM}.  For $P<1000$~d, $e$ was chosen from a Rayleigh distribution ($dp \propto e \exp(-\lambda e^2) de$) with $\langle e^2 \rangle^{1/2} = (2 \lambda)^{-1/2} = 0.33$; and for $P>1000$~d, $e$ was chosen from an Ambartsumian distribution ($dp = 2e de$), corresponding to a uniform distribution on the energy surface in phase space.  One might suppose that the observed eccentricity distribution is peaked towards small values for $P<1000$~d solely because of tidal dissipation, but this cannot be the case.  In particular, only one out of 12 isolated binaries with $11 \hbox{ d} < P< 1000 \hbox{ d}$ in the volume-limited sample of \cite{1991DM} has $e>0.5$.  However, tidal dissipation would only be able to affect the orbit of a main-sequence binary with $P \approx 100$~d if $e \gtrsim 0.8$.  Therefore, the small eccentricities of binaries with $P < 1000$~d cannot mostly result from the circularization of binaries with larger eccentricity.

A total of $7 \times 10^4$ such binaries were selected, and the evolution of $P$, $e$, $\Omega_1$ and $\Omega_2$ were followed using the equations of \S\ref{eom}.  The integrations proceed very quickly, since the spins were assumed to be aligned with the orbit---only the magnitudes of the vectors in equations (\ref{edoteq})-(\ref{om2doteq}) needed to be followed, not their orientation.  The integrations were stopped when $e \leq 10^{-3}$ or at 10~Gyr (roughly the main sequence lifetime of a $1 M_\sun$ star), whichever came first.

\begin{figure}
\includegraphics*[scale=0.85]{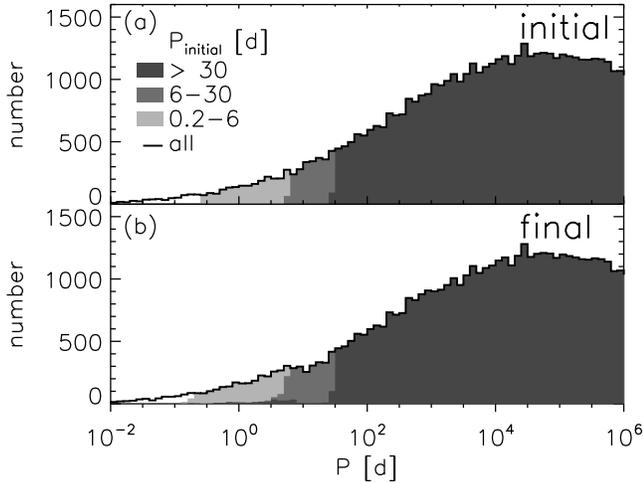}
\caption{The periods of isolated binaries before and after 10 Gyr of evolution by tidal friction.  (a) Histogram of the assumed initial period distribution.  (b) Histogram of the final period distribution, showing no dramatic change. \vspace{0.3 in} }
\label{period2body}
\end{figure}

In Figure~\ref{period2body} we plot the initial and final period distributions.  The Gaussian tail at long periods is not plotted.  Histograms corresponding to various initial period ranges are plotted in different shades.  In particular, if the primordial period distribution were cut off below 6~d, both the initial and final histograms would consist only of the two darker shades.  The unshaded region is probably not physical---main-sequence binaries with such periods would merge.  

A firm result is that tidal dissipation alone causes very little change to the period distribution of binaries.  We turn now to KCTF.

\begin{figure}
\includegraphics*[scale=0.85]{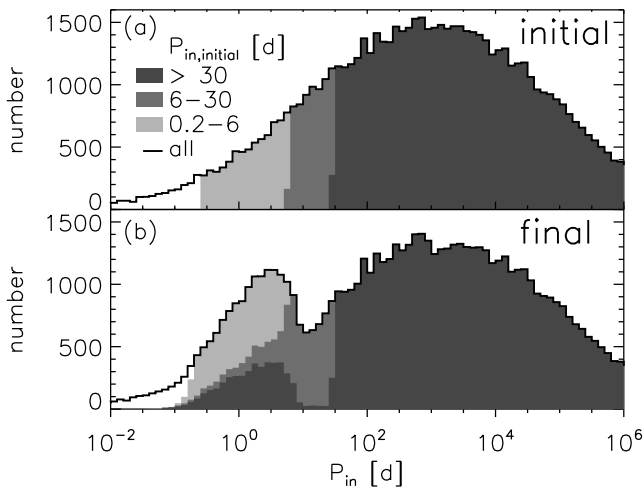}
\caption{The periods of the inner binaries of the simulated triples before and after 10 Gyr of evolution by Kozai cycles with tidal friction.  (a) Histogram of the assumed initial inner period distribution.  (b) Histogram of the final period distribution, showing the production of numerous close binaries with $0.1\hbox{ d} \lesssim P_{in,final} \lesssim 10\hbox{ d}$, many of which initially had much longer periods. \vspace{0.3 in} }
\label{period_dist}
\end{figure}

\section{Population of triples} \label{mc-calc}
In this section we calculate the semi-major axis and inclination distributions that result from KCTF.  We generated systems by Monte Carlo methods, then integrated their equations of motion (\S\ref{eom}).  

\subsection{Initial conditions}

The initial orbital distributions for both the inner and outer binaries were taken from the observed distributions of \cite{1991DM}.  Both the orbital distributions of the binaries and the physical parameters of the stars of the inner binary are chosen as above (\S\ref{sec:isobin}).  

For the tertiary, $m_3$ was determined by choosing $q_{out} = m_3/(m_1+m_2)$ from a Gaussian distribution as for $q_{in}$.  This approach implies that the mass of the third star was correlated with the mass of the inner binary, but we do not believe that this correlation has any significant effect on our results.  

Two periods and eccentricities were picked as above.  The smaller (larger) period was assigned to the inner (outer) orbit.  The semi-major axes were computed from these masses and periods assuming non-interacting Keplerian orbits.  The mutual inclination distribution of the tertiary is assumed to be isotropic with respect to the inner binary; thus we selected $\cos i$ to be uniform in $[-1, 1]$.  The other angles, $\omega_{in}$ and $\Omega_{in}$, were selected uniformly in $[0, 2 \pi]$.

After these parameters were selected, we used the empirical stability criterion of \cite{2001MA} to determine whether the system is hierarchical or if it will disrupt in a small number of dynamical times.  If the semi-major axes obeyed the criterion:
\begin{equation}
a_{out}/a_{in} > 2.8 (1 + q_{out})^{2/5} \frac{(1+e_{out})^{2/5}}{ (1-e_{out})^{6/5}} (1 - 0.3 i / 180^{\circ}), \label{stable_condition}
\end{equation}
then we accepted the triple as stable and integrated its averaged equations of motion.  Otherwise, we assumed it disrupted, resulting in an unbound binary and single star (we do not include those binaries in the following results).  About 40\% of selected triples failed to fulfill the condition (\ref{stable_condition}).  A total of $7 \times 10^4$ stable systems were integrated and the results are presented here.

\subsection{Stopping conditions}
In most cases we stopped the integrations at $10$~Gyr, roughly the main sequence lifetime of a $1 M_\sun$ star.  However, for some systems a straight-forward integration of the averaged equations of motion was prohibitively expensive.  In these cases we used the following procedure to deduce the final state without a costly integration.

The largest such group is triples whose Kozai cycle does not cause pericenter passages close enough for tidal dissipation to be effective.  For these we integrated the equations until the first eccentricity maximum and computed the eccentricity damping timescale $(V_1 + V_2)^{-1}$ (eq.~[\ref{eq:V}]) there.  If it was longer than $10$~Gyr, we integrated until a second eccentricity maximum, then took the properties at a random time in the interval between the maxima, similar to the method of \cite{2005TR}.  These systems will oscillate for their whole main sequence lifetimes, so choosing a random point of an oscillation near the initial time is statistically indistinguishable from a random time at the currently observed epoch.  

If the triple is strongly hierarchical initially, or if it is driven to such a state by KCTF, then the pericenter precession due to relativity and stellar distortion dominates that of the third body.  As shown in \S\ref{mod-ei}, Kozai cycles are suppressed in this case, so the subsequent evolution is very similar to that of an isolated binary.  Therefore we stopped the integration once the Kozai timescale was more than $30$ times the pericenter precession period, and evolved the system to $10$~Gyr by the equations for eccentricity damping neglecting the third body (see \S\ref{sec:isobin}).  In real systems, the third body actually continues to have a modest effect which is not modeled by the orbit-averaged equations of motion \citep{1979MS}, which we neglect.  As long as this timescale criterion was satisfied, we also neglect the effect that nodal precession of the inner orbit has on tidal dissipation.

Finally, some systems took many thousands of Kozai cycles to evolve significantly.  We stopped individual integrations at 4 CPU minutes, and if the system had not reached 10 Gyr and appeared still to be evolving, we re-integrated it with an artificially small viscous time so that the evolution would take place in $\sim100$ Kozai cycles.  For some individual systems we checked that the final state of this integration had parameters to within a percent of those of the final states of the original systems.  These systems either reached the end of their allotted time, which was scaled down from 10 Gyr in proportion to the scaling of the viscous time, or stopped oscillating, which allowed the neglect of the third body in integrating the further evolution as above.

\subsection{Numerical results}
Figure~\ref{period_dist} shows the relation between the initial and final period distributions for the inner binary.  Shaded portions of the histogram show how initial periods map to final periods.  The main result is the strong peak in the distribution of periods near $P_{in,final} \simeq 3$~d.  The great majority of these systems have evolved onto circular inner binary orbits for which the perturbation of the third star no longer causes interesting effects.  As an aside, we note that the initial period distribution of inner binaries (Fig.~\ref{period_dist}a) is peaked towards lower values than the initial period distribution of isolated binaries displayed in Figure~\ref{period2body}a; the latter is simply the \cite{1991DM} distribution.  This difference is a consequence of the definition of an inner binary, which biases inner binaries to lower periods and outer binaries to higher periods.

\begin{figure}
\includegraphics*[scale=0.85]{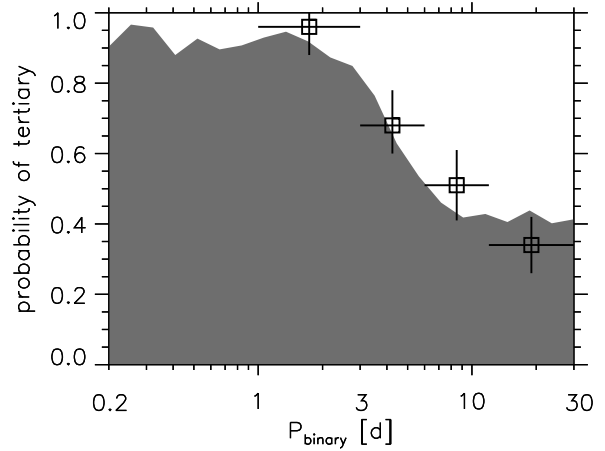}
\caption{ The fraction of binaries with tertiaries.  The points are from the observational study of \cite{2006T}---horizontal bars indicate the period range and vertical bars represent the error on the fraction of tertiaries.   The gray theoretical histogram is constructed via a linear combination of the final distributions computed in \S\ref{sec:isobin} and \S\ref{mc-calc}.  Two free parameters were varied to achieve best fit with the observational results: (a) the overall fraction of triples relative to all systems (binaries plus triples)---0.25 was best-fit---and (b) the cutoff period of the primordial distribution---6 d was best-fit, corresponding to histograms in Figures~\ref{period2body}b and \ref{period_dist}b including only the two darkest shades. \vspace{0.3 in}  }
\label{compare_tokov}
\end{figure}

Let us pause to compare the final period distribution to the observed systems.  In Figure~\ref{compare_tokov} we plot the fraction of spectroscopic binaries determined to have a tertiary by \cite{2006T}.  We also plot a theoretical distribution determined by our integrations.  This distribution was constructed by taking a linear combination of the final period histograms for isolated binaries (Fig.~\ref{period2body}b) and for inner binaries of triples (Fig.~\ref{period_dist}b).  Two free parameters were determined by fitting the data from \cite{2006T}: (a) the relative number of triple systems to all systems---both binaries and triples---and (b) the primordial cutoff to the period distributions.  The best-fitting proportion of triple systems to all systems is $0.25$, and the best-fitting primordial cutoff period is $6$~d.

These parameters are consistent with independent estimates.  The observations of additional components to binary stars are quite incomplete, but approximately one-third of visual binary stars have an additional component \citep{2002TS, 2004T}, consistent with our estimate.  A complication is that the observed mutual inclination distribution of triple stars with long-period inner binaries shows moderate correlation between the inner and outer orbits \citep{1993T}.  Recent data suggests $\langle i \rangle \approx 73^\circ \pm 6^\circ$ \citep{2002ST}, and the distribution can be reasonably represented by the sum of an isotropic distribution ($\cos i$ uniform in $[-1, 1]$; 75\% of systems) and a strongly correlated distribution ($i$ uniform in $[0^\circ, 40^\circ]$; 25\% of systems).  The former are the systems we have simulated, and the latter do not evolve substantially by KCTF.  Therefore, due to this correlation, our overall tertiary fraction of $0.25$ may need to be augmented by a factor of $\sim1/3$, in which case it is still consistent with the triple frequency estimated by Tokovinin.  
Similarly, our best-fitting period cutoff is consistent with the size of protostars, once they become dynamically stable, only contracting on the Kelvin-Helmholtz timescale.  For example, a solar mass star at an age of $7 \times 10^4$~yr has a radius of $\sim 6 R_\odot$ \citep{1994DM}; a binary consisting of two such stars in marginal contact would have a period of $3.4$~d.  

The comparison with the observed frequency of tertiaries as a function of period is an encouraging confirmation of the importance of KCTF for modifying the period distribution of binaries.  We turn now to predictions of the mutual inclination distribution, which can be tested by future observations.  

\begin{figure}
\includegraphics*[scale=0.9]{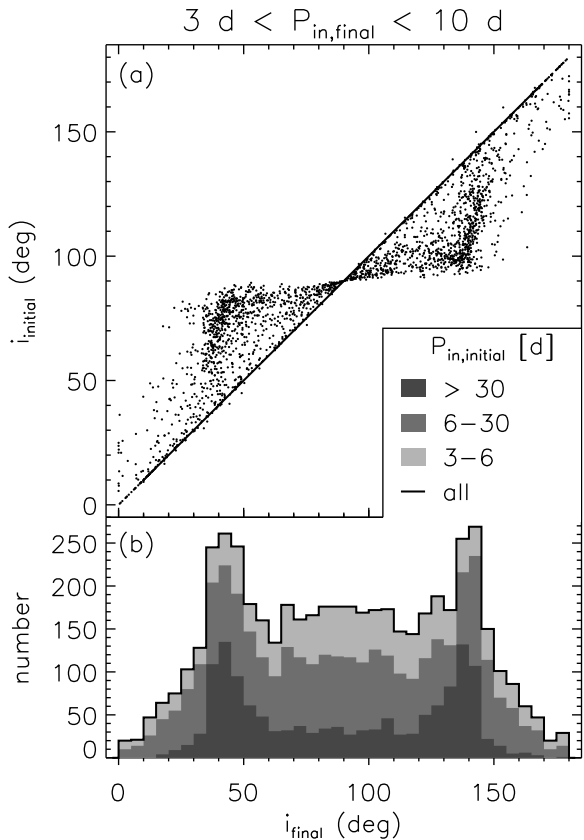}
\caption{ (a) The initial and final mutual inclinations for inner binaries with final periods between 3~d and 10~d.  Systems move away from $90^\circ$ during the high eccentricity phase of a Kozai cycle, so tidal dissipation during that phase tends to lock in a more moderate mutual inclination.  (b) Histogram of final mutual inclination for these systems.  The strong spikes correspond to the most probable mutual inclinations at the high eccentricity portion of Kozai cycles (compare to Fig.~\ref{contoureoi}b).  These spikes are strongest for binaries that have shrunk dramatically by Kozai cycles; e.g., if primordial binaries with $P_{in,initial} < 6$~d are rare, about twice as many systems per $5^\circ$ increment of $i$ are expected to inhabit the spikes relative to other configurations. \vspace{0.3 in}  }
\label{inclination_dist}
\end{figure}

The relationship between initial and final inclinations is illustrated in Figure~\ref{inclination_dist}a for systems with $P_{in,final}$ between 3 and 10~d.  During Kozai cycles, the inclination tends to move away from $90^\circ$.  Tidal dissipation at maximum eccentricity seals in these more moderate inclinations.  The most interesting feature in the distribution of $i_{final}$ (Fig.~\ref{inclination_dist}b) is the spikes that appear near $i_c = 39.2^\circ$ and $140.8^\circ$.  The shaded histograms show the result broken down by initial inner binary period.  Triple systems that start with $P_{in,initial}$ between 3~d and 10~d, but do not evolve by KCTF, contribute a rather isotropic distribution (lightest shading).  Therefore the spikes are fractionally much stronger if the primordial inner binary period distribution is cut off at a large period.  This inclination distribution is a distinctive feature of Kozai cycles, as predicted by the simple model of \S\ref{analsect} (see Fig.~\ref{contoureoi}b), and therefore would provide unambiguous observational evidence for KCTF.

In a small fraction of triples, the third star is close enough ($\lesssim 1$~AU) so that its strong perturbation causes alignment of the inner and outer binaries even after the inner binary's orbit has circularized.  Such systems generally evolve to either $i_{final} = 0^\circ$ or $i_{final} = 180^\circ$ (see Fig.~\ref{inclination_dist}a, but the majority of systems that undergo this process have $P_{in,final} < 3$~d).  The physical mechanism is that the spins of the stars of the inner binary cannot come into alignment with the orbit normal, because of nodal precession due to the third body.  Therefore tidal dissipation continues, draining energy from the inner orbit.  Here $H_{in}$ is still approximately conserved (only a small amount of angular momentum is transfered to the spins), so orbit shrinkage requires that the mutual inclination decreases.  See \cite{2007FJG} for a detailed description of this process.  If coplanar triple systems are found with $i = 0^\circ$, they might also be interpreted as resulting from fragmentation of a rather thin disk, but an observation of the purely retrograde case ($i = 180^\circ$) would seem to require this dissipation mechanism.  Our simulations yield equal numbers of systems with $i_{final}\simeq0^\circ$ and $i_{final}\simeq 180^\circ$, but this is an artifact of our assumption of isotropic initial conditions, which is probably not valid for such compact systems.

\begin{figure}
\includegraphics*[scale=0.85]{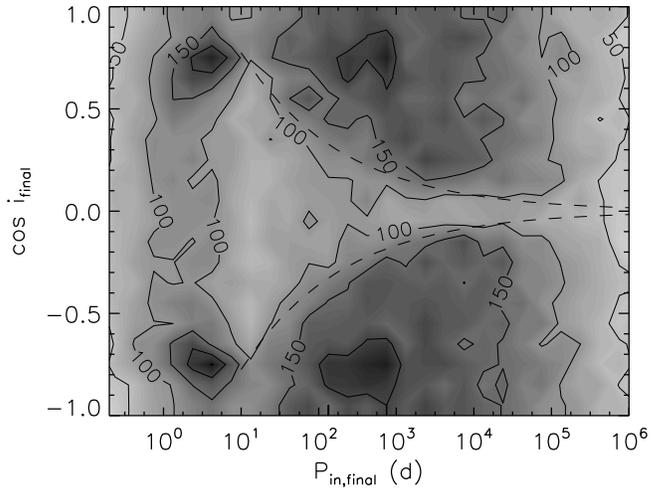}
\caption{Two dimensional histogram of the cosine of the final mutual inclination of inner and outer binaries versus the final inner binary period.  Contours are spaced in increments of 50, and the gray-scale represents the number density with a finer gradation.  Bin size is a quarter of an order of magnitude in $P_{in,final}$ and $0.1$ in $\cos i_{final}$, which is the same resolution as the tick marks.  The striking paucity of systems with $i_{final} \approx 90^\circ$ and $10 \hbox{ d} < P_{in,final} < 10^5\hbox{ d}$ is due to attrition: systems that started with these values have been removed by KCTF to smaller periods.  Many of these systems inhabit the spikes of Figure~\ref{inclination_dist}b, which are clear enhancements in this plot.  The region between the dashed lines corresponds to initial values of systems that may lose enough angular momentum by Kozai cycles, and enough orbital energy by tidal friction, to circularize to final periods less than 10~d (see text).  \vspace{0.3 in} }
\label{fig:pvsi}
\end{figure}

Now we step back from these small periods to survey the whole range of periods affected by KCTF.  In Figure~\ref{fig:pvsi} we plot the distribution of systems as a function of both $\cos i_{final}$ and $P_{in,final}$.  The spikes of Figure~\ref{inclination_dist}b are prominent at low periods, but there is another striking feature: a deficit of long-period inner binaries with $i \approx 90^\circ$.  These systems have evolved to lower periods by KCTF.  Here it is clear that KCTF can remove a significant fraction of near-perpendicular systems to small periods, over a wide range of inner binary periods, from $\sim 10$~d to $\sim 10^5$~d.  For comparison, consider an initially circular inner binary that undergoes eccentricity growth by the Kozai mechanism, then is tidally circularized from $e_{in,max}$ at constant orbital angular momentum.  By this prescription, the locus of initial systems that produce binaries with $P_{in,final} = 10$~d is indicated by dashed lines on Figure~\ref{fig:pvsi}.  Systems between those lines may be expected to evolve significantly by KCTF, if other sources of pericenter precession are negligible.  The deficit of triple systems with perpendicular orbits was identified by \cite{1968H} in the first theoretical paper on KCTF, and this effect should be tested by determining the mutual inclination distribution of triple stars.  

\section{Persistence of the final mutual inclination} \label{finali}
The final product of KCTF is a triple system with a very large period ratio $P_{out}/P_{in}$, a circular inner orbit, and a mutual inclination that is often near the critical one.  Since the distinctive inclination distribution that results from KCTF (Fig.~\ref{inclination_dist}b) may ultimately provide the strongest observational evidence for this process, it is important to ask whether the inclination persists from the end of KCTF to the present.  We point out three mechanisms for changing the mutual inclination but argue that each has a negligible effect.  

First, the Galactic tide may cause the outer binary to precess. The
precession period for a binary with period $P_b$ is roughly
$P_{gal}^2/P_b$, where $P_{gal}$ is the vertical oscillation period of
the star in the Galactic disk, with a local value of $[\pi / (G \rho) ]^{1/2} \approx 10^8 {\rm yr} [\rho / (0.1 M_\sun {\rm pc}^{-3})]^{1/2}$. For the precession period of the outer binary to be
less than the age of the Galaxy, $P_0\approx 10$~Gyr, we require $P_b
\gtrsim P_{gal}^2/P_0 \approx 1$~Myr.  That is, the Galactic tide can change the orientation of the outer binary substantially only if $a_{out} \gtrsim 10^4$~AU.  Pericenter precession of the inner binary due to such a distant third
body will likely to be overwhelmed by relativistic precession;
equation~(\ref{emax_gr}) implies that
$\tau\dot\omega_{GR}|_{e_{in}=0}$ must be $\lesssim 1$ for
substantial eccentricity oscillations, so Kozai cycles will be
suppressed unless $a_{in,initial}$ is very large.  For example, an
inner binary with an initially circular orbit can only satisfy this
constraint if $a_{in,initial} \gtrsim 25$~AU, given $a_{out} = 10^4$
AU.  For an orbit with $a_{in,initial} = 75$~AU, the pericenter
distance $\lesssim 0.03$~AU required for substantial tidal friction
requires a maximum eccentricity in the Kozai cycle given by
$1-e_{max}< 0.0004$, and this can only be achieved if the orbits are
initially almost exactly perpendicular ($|i-90^\circ| \lesssim
1^\circ$). Larger or smaller values of $a_{in,initial}$ have an even stricter
requirement.  Therefore the number of systems for which KCTF yields a
close binary, after which the Galactic tide reorients the outer
binary, is negligibly small.

Second, individual passing stars will perturb the outer orbit, changing its orbital elements.  The timescale for changing the outer binary's angular momentum is similar to the timescale of changing its energy, which corresponds to the disruption timescale.  For the typical density ($0.1 M_{\sun} /{\rm pc}^3$) and velocity dispersion ($\sim 40$~km/s) of stars in the disk, this timescale is shorter than $10$~Gyr only for binary semi-major axes above $\sim 3 \times 10^4$~AU.  As above, such distant third bodies will not produce short-period inner binaries via KCTF with any reasonable frequency, so mutual inclination change by passing stars is also ineffective.

Finally, let us specialize to a planet that has migrated by KCTF.  Generally its orbit will not lie in the host star's equatorial plane, since tidal dissipation in the star will be too weak to align the stellar spin with the planetary orbit (see \S\ref{RMeffect}).  Therefore, the planet induces a spin precession of its host star with a period of $\sim10^5$~yrs.  As the stellar spin precesses, the planet's orbit must also precess to conserve angular momentum.  However, the magnitude of the associated change in $i$ is $\lesssim 10^\circ$, since the orbit of a hot Jupiter usually has more angular momentum than the spin of its host, so this effect is unlikely to change the inclination distribution significantly.  The thick appearance of the line in the final state in Figure~\ref{kctf_wm}c is this type of oscillation.

\section{Application to hot Jupiters} \label{hotjup}

Stars are born by the fragmentation of molecular clouds.  Planets, however, are believed to form in protostellar accretion disks, after material has settled around a star.  In binary star systems, it might be expected that the angular momentum of the disk around each star would come into alignment with the angular momentum of the binary orbit.  The star accretes high angular momentum material from this orbit-aligned disk, so the spin of the host star of the planetary system would likely be in alignment with the companion orbit as well.  Thus we might expect that both stellar spin angular momenta, the stellar orbital angular momentum about the system barycenter, and the planetary orbital angular momentum around its host star are all aligned if the binary semi-major axis is not too large.  \cite{1994H} has measured inclination to the line of sight of the spins of stars in binaries by comparing the rotational period of starspots to the $v \sin i$ values of rotationally-broadened lines.  If the stellar spins are aligned with each other, they will have zero difference in inclination to the line of sight.  The converse is not true, but the prevalence of alignment can still be assessed statistically.  Thus \cite{1994H} inferred that binaries are spin-aligned for $a \lesssim 30-40$~AU, but become randomly oriented for larger orbits.  Therefore, for nearly the entire exoplanet sample in binaries, it is likely that the protoplanetary disk was not aligned with the companion's orbit.  Therefore planets arising from these non-aligned disks may have a high inclination with respect to the binary companion, so KCTF may cause substantial evolution to the orbital distributions of planets in binary systems.  An
 alternative, but rarer, mechanism to produce non-aligned planets in
 binary star systems is to form a planetary system around a single
 star, to which is later added a binary companion through a dynamical
 interaction in the birth cluster \citep{2006PM}.

In the past decade, about 200 giant planets have been discovered, of which about 20\% have orbital periods short enough so that tides are important.  The high frequency of close-in planets was a surprise, since planet formation theory suggested that giant planets can only form beyond several AU, consistent with the current structure of the solar system.  Therefore, a migration mechanism is needed to reduce the angular momentum of these planets by a factor of $\sim10$.  The leading candidate is disk migration, in which torques between
 the planet and the remnant protoplanetary nebula transfer angular momentum from the planet to the gas \citep{1980GT, 1997W}.  There is some statistical evidence that planets orbiting close to one member of a wide binary have different
 properties---and hence a different formation or migration
 history---from planets orbiting isolated stars.  Among short-period radial velocity planets ($P<100$~d), the most massive planets ($M_{p,min} > 2$~Jupiter masses) are preferentially found in binaries \citep{2002ZM}, and these binary companions are close enough ($a_{out} \lesssim 300$~AU) that KCTF may operate \citep{2007DB}.  These short-period, massive planets in binary stars also have lower eccentricities than short-period planets orbiting single stars \citep{2004EUM}.  For these systems, KCTF provides an alternative migration mechanism to interactions with the
 protoplanetary nebula.  The predictions of the previous
 sections for the period and mutual inclination distributions should
 still apply in the planetary case, and there is an additional
 prediction of KCTF regarding the alignment of the orbital plane with the stellar equatorial plane (see \S\ref{RMeffect} below).

\subsection{Extra precession due to other planets}

In the solar system, apsidal precession is dominated by the gravitational perturbations from other planets; in multi-planet systems that are hosted by one member of a binary star, precession from other planets will compete with the precession due to the companion star \citep{1997HTT}.  \cite{2003WM} showed that the Kozai mechanism could be suppressed by small, undetected masses in the HD 80606 system, since its companion star is distant so the tidal field and precession rate from it are small.  \cite{1997I} have investigated the instability of the solar system's giant planets that would be caused by a companion star in an inclined orbit, as a function of inclination and companion mass.  \cite{2005M} have investigated how a close ($\sim 50$~AU) binary companion, coplanar with the planets, affects their scattering.  The Kozai mechanism, in systems in which it operates, may also lead to planet-planet scattering \citep{2006MDC}.

The key to why massive, short-period planets are preferentially found in binaries may lie in the competition between mutual planetary precession and precession due to the companion star, but this is a complicated process and the sign of the predictions is not clear.

Take, as an example, an ensemble of planetary systems in which the total mass of the planets is always the same, but the mass assigned to individual planets varies randomly.  Consider a case in which one planet is substantially more massive than the others.  Initially the mutual perturbations of the planets suppress Kozai oscillations in all of them.  Once moderate eccentricities and inclinations are built up in the planetary system, perhaps by mutual perturbations rather than the Kozai mechanism, the small planets become destabilized and are ejected.  These ejected bodies no longer suppress Kozai cycles, so now KCTF may cause migration for the largest planet.  Since the Kozai cycles depended on the planet ejecting its neighbors, this argument suggests that KCTF migration is more effective for more massive planets.

As a foil, consider an ensemble of planetary systems in which the total mass of the planets varies randomly, but is always divided into some number $N$ of planets of equal mass. If the total planetary mass is small enough, Kozai cycles will proceed unhindered, and one or more planets may migrate by KCTF.  The presence of the small planets that failed to suppress Kozai cycles may not be detectable with radial velocity surveys, but the planet that migrated may be detected due to the larger reflex velocity that it induces in the star. This argument suggests that KCTF migration is more effective for less massive planets.

Since these two arguments yield opposite conclusions, at this time we can make no prediction regarding whether KCTF migration favors high or low mass giant planets. 

\subsection{The period distribution of planets in binaries} \label{sec:planetperiods}

\begin{figure}
\plotone{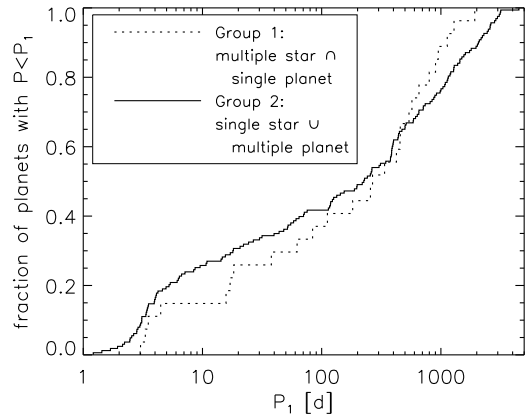}
\caption{ The cumulative distributions of two populations of planets: those for which Kozai cycles may have been possible (Group 1) and those for which they are not (Group 2).  Group 1
 consists of all extrasolar planets having a binary stellar companion
 but no other detected planets; Group 2 consists of all other known
 extrasolar planets.  There is no statistically significant difference in the distributions, thus no statistical evidence for KCTF in the period distribution of planets.  In contrast, to date the strongest evidence of KCTF in triple stars is the period distribution of inner binaries \citep{2006T}. \vspace{0.3 in}  }
\label{pcum}
\end{figure}

The period distribution produced by KCTF (Figure~\ref{period_dist}) is quite similar to the ``pile-up'' of hot Jupiters near $3$~d \citep{2006FR}.   We have conducted the following statistical test to look for direct evidence of KCTF in the period distribution of the
 exoplanet sample.  We split the radial velocity planets\footnote{Downloaded on September 15, 2006 from \url{http://vo.obspm.fr/exoplanetes/encyclo/catalog.php} } into two groups: (1) single planets in a system with multiple stars according to \cite{2006R}, and (2) planets orbiting single stars and planets in multiple-planet systems.  Group (1) contains the systems that could exhibit Kozai cycles, now or in the past, since a third body is known and there are not other planets that could suppress Kozai cycles.  Group (2) are the planets for which Kozai cycles are less likely to be present (assuming that multiple planetary systems are roughly coplanar and that they suppress Kozai cycles driven by a stellar companion).  There may be systems with single planets and an \emph{undetected} third body that causes Kozai cycles \citep{2005TR}, but this possibility
 simply reduces the power---and should not bias---the statistical
 test.  In Figure~\ref{pcum} we plot the cumulative distributions of planetary period for these two groups.  We employ a Kolomogorov-Smirnov test to assess whether they are statistically indistinguishable.  The biggest gap between the distributions lies at $P=15.766$~d, actually in the direction opposite from the expectation of KCTF; it is $D_n = 0.140$.  For $n_1 = 27$ and $n_2 = 163$, the probability of having a gap at least this large is $72\%$ \citep{1992P}.  Therefore, there is no statistical evidence for KCTF in the period distribution of exoplanets.  As noted earlier, \cite{2006T} have employed the same test, finding different period distributions for isolated binaries and binaries in triples with a probability of $0.999$, which supports KCTF in triple star systems.

\subsection{Planetary inflation}

The energy associated with a hot Jupiter's orbit is about 10 times its binding energy.  If the planet has a high eccentricity and small pericenter distance, then its radius will inflate from tidal heating as the orbit circularizes \citep{2001BLM, 2003GLB, 2003BLL}.  For instance, for HD 80606b, which may be undergoing KCTF migration currently, \cite{2003WM} estimate a tidal luminosity of $10^{28}$~erg~s$^{-1}$.  For its minimum mass of $3.4$ times that of Jupiter, energy deposition at this rate would inflate the radius by about $30\%$ \citep{2001BLM}.  The increased radius will shorten the circularization timescale by a factor of $\sim 9$, so the planetary radius should be included in self-consistent integrations of KCTF migration.  Since the timescale for circularizing is a very strong function of the radius of the planet, the maximum semi-major axis out to which planetary orbits are
 circularized will be diagnostic of the planetary radius at migration.  There are three massive ($>2$~Jupiter mass) exoplanets in binaries whose orbits are circular, two of which (Gl 86b and HD 195019b) have larger periods ($15.8$~d and $18.2$~d, respectively) than exoplanets with circular orbits about single stars \citep{2002ZM, 2004EUM}.  \cite{2007DB} argue that tidal circularization is ineffective at these large periods, but we suggest that this observation, if
 confirmed in a larger sample, not only supports the KCTF hypothesis
 but also may constrain the amount of inflation that is produced
 during KCTF migration.  For this explanation to hold, the circularization timescale must remain shorter than the gravitational contraction timescale as the eccentricity tends towards zero.

In the planetary case, the lower limit for the pericenter distance resulting from Kozai cycles is set by Roche lobe overflow.  If a highly
 eccentric planetary orbit is circularized, conservation of angular
 momentum implies that the final radius of the circular orbit is
 twice the initial pericenter; thus we expect that the minimum radius
 of circular orbits produced by KCTF is twice the Roche limit (this
 conclusion is supported by hydrodynamic simulations; see \citealt{2005FRW}).  Indeed, the observed semi-major axis distribution of hot Jupiters appears to cut off at twice the Roche limit \citep{2006FR}.
 
This simple argument does not account for the tidal inflation instability of \cite{2003GLB}.  The planet may inflate enough to overflow its Roche lobe, then lose mass and increase its semi-major axis.  According to \cite{2003GLB}, starting from a large semi-major axis before inflation (which is natural for migration by KCTF), the planet will not be disrupted but mass loss will eventually halt, leaving the planetary orbit with a period of 1-3 weeks and possibly with residual eccentricity.  Some of the mass lost from the
 planet may be accreted onto the host star, leading to pollution of
 the star's photosphere with metals.  Two stars that form a binary are likely to have the same bulk metallicity, so enhanced photospheric metallicity in the star hosting a hot Jupiter is evidence for accretion of planet material after the pre-main sequence stage.  Furthermore, if isotopes that are known to be destroyed in stars (e.g., $^6$Li) are found in large abundance, a constraint on when the planetary material accreted can be given \citep{2001IMR}.

These effects may be tested for the planet HAT-P-1b \citep{2007B}, a transiting planet in a binary stellar system.  It could be in the final stages of circularization after KCTF migration, which would lead both to some residual eccentricity and to an inflated radius; the latter is securely observed.

\subsection{Spin-orbit alignment} \label{RMeffect}

In close binary star systems, including those brought close by KCTF, the spins of the stars are expected to align with the orbital angular momentum in a timescale short compared to the circularization timescale.  However, a star's spin probably does not align with the orbital angular momentum of a hot Jupiter it hosts because of the small mass of the planet (see Table~\ref{table:spins}).  A measurement of $\psi$, the angle between the stellar spin angular momentum and the planet orbital momentum, can therefore constrain the history of the system.

\begin{table}
\caption{Expected alignment properties of most triples after KCTF.}
\begin{tabular}[b]{c | cc}
 & \multicolumn{2}{c}{system type: $m_1$-$m_2$-$m_3$} \\
alignment of: & star-star-star & star-planet-star \\
\hline
$m_2$-spin / inner-orbit & yes & yes \\
$m_1$-spin / inner-orbit & yes & no \\
inner-orbit / outer-orbit & \multicolumn{2}{c}{no; $i \approx 40^{\circ}$ or $140^{\circ}$ likely} \\
\hline
\end{tabular}
\label{table:spins}
\end{table}

It can be expected that if a planet migrated via interactions with the disk, it will remain in the same orbital plane to within an angle much less than a radian.  However, for KCTF migration, the angular momentum of the planet precesses about the angular momentum of the inclined companion.  This precession happens generally when a body in an inclined orbit is introduced.  Even the coplanarity of the solar system (whose net orbital angular momentum is dominated by Jupiter, and $\psi = 7^\circ$) provides constraints on unseen planets in inclined orbits at large distances \citep{1972GW}.  As tidal dissipation takes over during KCTF migration, the planet's orbit eventually couples more strongly to its host's equatorial bulge than to the stellar companion, and misalignment persists as a fossil record of the earlier period of precession driven by the companion.  Therefore, if planets are formed and migrate via disk torques to become hot Jupiters on a timescale short compared to the precession timescale due to the stellar companion, then we may expect to see approximate spin-orbital alignment even in binary systems.  On the other hand, if there was a period in which the planet's orbital evolution was dominated by the binary companion, and later it came to be a hot Jupiter through KCTF migration, then non-alignment between stellar spin and planetary orbit will be the norm.

We integrated a series of 1000 systems to illustrate this effect.  Each system had identical parameters except for $i_{initial}$, which ranged from $84.3^\circ$ to $90^\circ$ (an evenly-spaced grid of $\cos i_{initial}$ between $0$ and $0.1$).  Every system had $m_1 = m_3 = 1 M_\odot$, $m_2 = 10^{-3} M_\odot$, and started with $a_{in}=5$~AU, $e_{in}=0.1$, and $\omega_{in}=\Omega_{in}=0^\circ$.  The outer binary had $a_{out}=500$~AU and $e_{out}=0$.  Before the planet has migrated at all, the Kozai timescale for these systems is $\tau_{initial}=2.36$~Myr (eq.~[\ref{tau}]).  The host star's spin period was set to 10 d, beginning with zero obliquity ($\psi = 0$).  The planet was given a viscous time $t_V = 0.01$~yr (see appendix) and started with no spin.  Once the orbit has shrunk to $a_{in}=0.15$~AU, planetary orbital plane precession is dominated by the host star's bulge rather than the companion, so $\psi$ stops evolving; we call the time it takes to reach this point $t_{a=0.15 AU}$.  The system with the smallest inclination migrated the slowest, with $t_{a=0.15 AU} = 4.5$~Gyr.  We found that a retrograde system with $i_{initial} = 90^\circ + \delta i$ has the same value of $\psi_{final}$ as a system with $i_{initial} = 90^\circ - \delta i$, which is required by the symmetries of the equations of motion and the chosen initial conditions.  Thus, the final distribution of $\psi$ from this calculation is a prediction for hot Jupiters that have migrated by KCTF after beginning with isotropic inclination relative to a companion star.

\begin{figure}
\includegraphics*[scale=0.85]{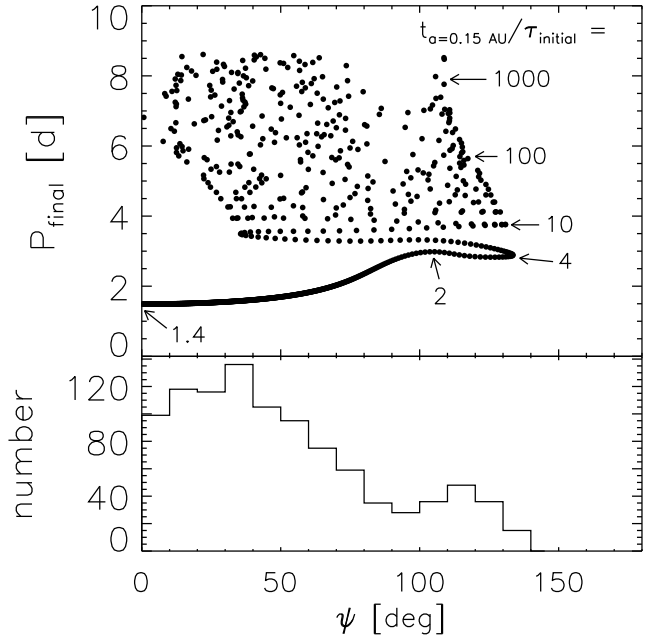}
\caption{ (Top) The final period of planets after KCTF migration versus the final stellar obliquity $\psi$ (the angle between stellar spin and planetary orbit)---see \S\ref{RMeffect} for initial conditions.  These integrations neglected planetary inflation, which we have argued may affect the final periods.  For final periods less than $\sim 2.5$~d, the planetary orbit reached only one eccentricity maximum, so it precessed less than one radian due to the stellar companion.  Certain planets in the sequence are labeled by the length of time they took to migrate from $a_{in}=5$~AU to $0.15$~AU, in units of the initial Kozai timescale $\tau=2.36$~Myr (eq.~[\ref{tau}]).  (Bottom)  Histogram of $\psi$.  Misalignment is common, and even retrograde ($\psi > 90^\circ$) orbits are possible.  This misalignment is observable in transiting planets through the Rossiter-McLaughlin effect and would provide strong evidence for KCTF migration of a hot Jupiter in a wide binary system. \vspace{0.3 in} }
\label{fig:spinorbit}
\end{figure}

The results are given in Figure~\ref{fig:spinorbit}.  Systems with $i_{initial}$ within $3.3^\circ$ of $90^\circ$, corresponding to 60\% of the total, only undergo one Kozai cycle; i.e., $e_{in}$ attains a single maximum, after which $a_{in}$ and $e_{in}$ are damped.  Such systems precess a fraction of a radian, therefore they form an orderly sequence at the smallest periods ($P_{final} < 2.5$~d).  Systems that finished at longer periods underwent many Kozai cycles and the orbital plane sampled many orientations.  For such systems, two aspects of the evolution allow us to predict the range of $\psi$ attained.  First, during most of these cycles, the host star maintains its space orientation, as the torque from the planetary orbit is small.  Second, tidal dissipation is only important when $e_{in}$ is near its maximum, such that $i$ is close to $i_{min} \approx 40^\circ$ (Fig.~\ref{contoureoi}).  Together, these considerations imply the systems will produce hot Jupiters whose host stars have obliquities in the range $i_{initial} - i_{min} \lesssim \psi \lesssim i_{initial} + i_{min}$.  This range approximately describes the results for planets with $P_{final}$ between 3~d and 5~d.  For $P_{final} \gtrsim 5$~d the stellar spin is torqued through a large angle as the planet's eccentricity damps, partially realigning it with the planet's orbital angular momentum, sliding the range of final $\psi$ to lower values.  (We have verified that this final partial realignment does not occur if the planet's mass is $10^{-6} M_\odot$; its orbital angular momentum is then only $\sim 1\%$ of the stellar spin momentum, so the planet cannot substantially reorient the star.)

So far, besides the solar system, there are only four planetary systems for which $\psi$ has been constrained (see Table~\ref{table:alignment}).   The Rossiter-McLaughlin (RM) effect \citep{1924R, 1924M, 2007GW} constrains the sky-projection of $\psi$ (called $\lambda$ in the literature) for planets that transit their host star.  The RM effect arises because the spectral lines of the star are rotationally broadened, and the planet blocks portions that are red-shifted or blue-shifted as a function of time when it transits the stellar disk.  The derived values of the alignment angle are effectively sky-projected (hence underestimated), but they are of the same magnitude as the solar system and are inconsistent with the distribution of $\psi$ in Figure~\ref{fig:spinorbit}, suggesting that at most a modest fraction of hot Jupiters are produced by KCTF.  A prime candidate for the measurement of the RM effect is the transiting planet HAT-P-1b \citep{2007B}, for which misalignment would be corroborating evidence for the KCTF migration hypothesis, made plausible by its inflated radius.

\begin{table}
\caption{Measurements of the sky-projection of the angle between stellar spin and planetary orbit ($\lambda$) for transiting exoplanets.  Of these, only HD 189733 is known to be a stellar binary. }
\begin{tabular}{c  c c}
Exoplanet & $\lambda$ & Reference  \\
HD 209458b & $-4.4^\circ \pm 1.4^\circ$ & \cite{2005Wa} \\
HD 189733b & $-1.4^\circ \pm 1.1^\circ$ & \cite{2006W}\\
HD 149026b & $11^\circ \pm 14^\circ$ & \cite{2006Wo}\\
TrES-1b & $30^\circ \pm 21^\circ$ & \cite{2007N} \end{tabular}
\label{table:alignment}
\end{table}

Alternatively, the timing of migration can be strongly constrained by the precise alignment of the transiting planet HD 189733b \citep{2006W}.  According to \cite{2006Ba} there is an M-dwarf companion at a projected distance of 219 AU, and preliminary measurements exclude coplanarity between the inner and outer orbits at the $4-\sigma$ level.  If the outer orbit's misalignment is confirmed, it may be possible to constrain the timescale of formation and disk migration to less than about $\tau \approx 2$~Myr, assuming the planet formed with a period of about $10$~yr and migrated before torque from the companion caused misalignment. 

Further, the precession of a planet's orbital plane due to its host star's rotational bulge may be detectable through a changing transit duration \citep{2002M}.  Large misalignment, which is common according to Figure~\ref{fig:spinorbit}, causes a large observable effect.

\section{Discussion} \label{sec:discuss}
We have described the
 evolution and final orbital element distribution of binary stars
 resulting from the combined effects of the gravitational influence
 of a tertiary companion star and tidal friction. In particular, we
 show that Kozai cycles plus tidal friction (KCTF) can strongly
 enhance the number of binary stars with periods in the range 0.1 d
 to 10 d; briefly, the distant companion induces strong eccentricity
 oscillations (Kozai cycles) in systems in which the inner and outer
 binary orbital planes are nearly perpendicular, and these eccentric
 orbits are circularized near pericenter by tidal friction.
 
The logarithmic period distribution
 of inner binaries produced by KCTF shows a peak near 3~d and
 declines from 3~d to 10~d. These features are consistent with the empirical period distribution of inner binaries of hierarchical systems found by \cite{2006T}, and we confirm these authors' interpretation that KCTF is the cause of this feature.  Therefore KCTF appears to produce most close binaries.
  
The dominance of KCTF for producing close binaries may be confirmed by testing our prediction of a modified mutual inclination distribution.  We expect an enhancement of systems with mutual inclination close to the critical values of $39.2^\circ$ or $140.8^\circ$ if $P_{in}$ is in the range $3$-$10$~d.  Interferometric measurements have been determining the mutual inclination in triple stars with inner periods small enough for tides to be important \citep{2006M}, and with improvements in limiting magnitude and precision they will be able to measure enough systems to test this theoretical prediction.  Planetary inclination angles relative to a stellar companion will in some cases be measurable by \emph{SIM PlanetQuest}.  

As the observations improve, the present theory should be augmented to refine the predictions.  In particular, most of the systems in the spikes in the inclination distribution (Fig.~\ref{inclination_dist}b) have $2 < G_{out} / L_{in} < 10$, so our assumption that the outer binary dominates the angular momentum is only marginally valid.  A more accurate model for the known three-body gravitational dynamics may also have some effect on the period distribution produced by KCTF.  For instance, \cite{2000FKR} showed that including octupole terms in the interaction Hamiltonian causes Kozai cycles to be quasi-periodic, and some pericenter approaches can be much closer than those computed with quadrupole terms alone (their Fig. 5).  These
 closer approaches may lead to smaller final periods after
 circularization than we have predicted.  Another aspect of the present theory that could be improved is modeling dynamical tides as well as, or instead of, the equilibrium tide.  This more sophisticated tidal theory follows how the tidal force excites free modes of oscillation in each star and how these modes are damped.  It may be useful in the present application, in which high eccentricities are common and the timescale of pericenter passage is comparable to the periods of oscillation modes.  However, the natural first step to refine the present analysis is to improve the model of point-mass gravitational interactions (which are completely understood) before attempting to improve the model of tidal interactions (which are quite uncertain).

We have investigated the period distribution that is produced by KCTF alone and found that it preferentially produces detached binaries with $P_{in} \simeq 2$~d.  However, over a timescale of a few billion years, magnetic winds may be able to extract enough angular momentum from such a binary so that its components come into contact or even merge \citep{1995S}.  A combination of KCTF and magnetic braking could therefore increase the space density of {\it contact} binaries.  KCTF and magnetic winds have often been regarded as competing theories of angular momentum loss for close binaries.  However, both are simple additions of an extra star to well-grounded empirical relations: angular momentum loss by magnetic winds adds a binary companion to the spin-down process of single stars, and KCTF adds a third star to the circularization process of close binary stars.  Therefore we believe both mechanisms must be operating at some level, and their relative contribution to the orbital evolution of close binaries is an interesting subject for future work.

We have considered hierarchical triple systems in the field of the Galaxy, where the space density of stars is small enough that the triples are presumably dynamically isolated (see, for instance, \S\ref{finali}).  Of course, triples could exhibit KCTF while embedded in a cluster of stars, where the dynamics would be much richer.  The outer binary of a primordial triple would be susceptible to disruption
 or reorientation by, or exchange with, passing stars during KCTF.  Also, dynamical interactions may {\it add} a third body to primordial binaries (at least temporarily), in which case the uncorrelated orientations of inner and outer binaries will be above the critical inclination $\sim$~77\% of the time, so Kozai oscillations will be the norm.  Some of these effects have been captured by the integrations of \cite{2001MA}, who added tidal dissipation to an N-body simulation of a star cluster.  A prime target for theorists is to explain the period distribution of binaries in the young, nearby Hyades cluster.  Its period distribution (for all binaries, not just inner binaries of hierarchical systems) shows a secondary peak at a few days that has eluded explanation \citep{1991DM}.  According to \cite{1995S}, angular momentum loss by magnetic winds cannot account for this feature---it tends to reduce the number of binaries at such periods, whereas a mechanism is needed to pile binaries up there.  We suggest that KCTF may be that mechanism.  To evaluate this hypothesis quantitatively, the above effects would need to be taken into account, as well as the cluster's young age which implies that KCTF has not yet progressed to completion in all cases.  KCTF should first establish binaries with periods less than 3 d in only a few Kozai cycles, then slowly add longer-period binaries over tens to hundreds of Kozai cycles (as an example, see the top panel of Fig.~\ref{fig:spinorbit}).  This prediction may be tested by surveying clusters of different ages.

Although KCTF appears to produce most close binary stars, KCTF may produce some but not most hot Jupiters, for two reasons.  First, systems with one planet orbiting a star of a binary system (i.e., those for which Kozai cycles are likely) have a planetary period distribution that is statistically indistinguishable from other planetary systems---see \S\ref{sec:planetperiods} and Figure~\ref{pcum}.  Second, the limited information on spin-orbit alignment for the four stars in Table~\ref{table:alignment} suggests that most host spins and planetary orbits are aligned, contrary to the prediction of KCTF---see \S\ref{RMeffect} and Figure~\ref{fig:spinorbit}.  For an individual binary star system that hosts a transiting planet, however, large misalignment detected by the Rossiter-McLaughlin effect would constitute strong evidence for KCTF in that system. 

\cite{2007DB} have shown that massive, short-period planets also tend to have binary companions within a few hundred AU, but those authors do not favor KCTF migration because two of those planets have periods apparently too long for tidal dissipation to be important.  However, during the period of extreme eccentricity set up by Kozai cycles, the planet's radius would expand due to the heat generated by tides, enhancing tidal friction.  Perhaps this effect can circularize planetary orbits to greater periods.  To evaluate this suggestion theoretically, planetary radius inflation must be modeled during calculations of KCTF.  Another worthwhile theoretical investigation would be the the suppression of Kozai cycles by mutual planetary perturbations.

\section{Conclusions} \label{sec:conclude}

We conclude by reviewing the predictions of KCTF:
(1) Period distributions for hot Jupiters and close binaries in systems with stellar companions have a peak at a few days and a minimum at longer periods ($\sim 10$~d) for which tidal circularization becomes ineffective.
(2) For triples in which the inner binary has a period between $3$ and $10$~d, there is an enhancement of systems whose mutual inclination between inner and outer binaries is near $40^\circ$ and $140^\circ$; for triples with longer inner periods, there is a paucity of systems with nearly perpendicular orbits.
(3) Hot Jupiters resulting from KCTF migration generally have orbital angular momentum misaligned with the stellar spin axis by large angles, frequently even larger than $90^\circ$, and they may have orbits circularized to greater periods than for hot Jupiters orbiting single stars due to inflation of the planetary radius during migration.

The role of KCTF in the formation of close binary stars and hot
Jupiters can be tested by measurements of the following quantities in
carefully defined samples of stars: (1) the period distribution of
close binaries that have a tertiary component; (2) the mutual
inclination distribution of triples, in order to detect the
enhancement of systems near the critical angles (these measurements
will normally require interferometry); (3) the angle between stellar
spin angular momentum and planetary orbital angular momentum for
planets hosted by stellar binaries, using the RM effect, transit
timing, or determination of the orbit of the outer binary for
transiting planets; (4) differential measurements of metallicity in
stars in binary systems that host planets, to detect stellar pollution from
overflow of the Roche lobe of the planet; (5) the distributions of
mass, eccentricity, and period for hot Jupiters.

\acknowledgments
This research is supported by NASA award NNG04H44G to ST.  We thank Bohdan Paczy{\'n}ski for initiating and supporting the project with great enthusiasm and remarkable insight.  We benefited from discussions with P. Eggleton, M. Krumholz, A. Tokovinin, M. van Kerkwijk, and Y. Wu.

\appendix

\section{Terms for the equations of motion} \label{app:tides}

Here we give the functional forms of the coefficients in the differential equations~(\ref{edoteq}-\ref{om2doteq}).

\begin{eqnarray}
V_1 &=& \frac{9}{t_{F1}} \left[ \frac{1 + (15/4)e_{in}^2 + (15/8)e_{in}^4 + (5/64)e_{in}^6}{(1-e_{in}^2)^{13/2}} - \frac{11 \Omega_{1h}}{18 \dot{l}_{in}} \frac{1 + (3/2) e_{in}^2 + (1/8) e_{in}^4}{(1-e_{in}^2)^5} \right], \label{eq:V}\\
W_1 &=& \frac{1}{t_{F1}} \left[ \frac{1 + (15/2)e_{in}^2 + (45/8)e_{in}^4 + (5/16)e_{in}^6}{(1-e_{in}^2)^{13/2}} - \frac{\Omega_{1h}}{\dot{l}_{in}} \frac{1 + 3 e_{in}^2 + (3/8) e_{in}^4}{(1-e_{in}^2)^5} \right], \\
X_1 &=& -\frac{m_2 k_1 R_1^5}{\mu \dot{l}_{in} a_{in}^5} \frac{\Omega_{1h} \Omega_{1e}}{(1-e_{in}^2)^2}  - \frac{ \Omega_{1q} }{ 2 \dot{l}_{in} t_{F1}} \frac{1 + (9/2) e_{in}^2 + (5/8) e_{in}^4}{(1-e_{in}^2)^5}, \label{Xeq} \\
Y_1 &=& -\frac{m_2 k_1 R_1^5}{\mu \dot{l}_{in} a_{in}^5} \frac{\Omega_{1h} \Omega_{1q}}{(1-e_{in}^2)^2}  + \frac{ \Omega_{1e} }{ 2 \dot{l}_{in} t_{F1}} \frac{1 + (3/2) e_{in}^2 + (1/8) e_{in}^4}{(1-e_{in}^2)^5}, \label{Yeq} \\
Z_1 &=& \frac{m_2 k_1 R_1^5}{ \mu \dot{l}_{in} a_{in}^5} \left[ \frac{2 \Omega_{1h}^2 - \Omega_{1q}^2 - \Omega_{1q}^2 }{ 2 (1-e_{in}^2)^2}  + \frac{ 15 G m_2}{ a_{in}^3 } \frac{1 + (3/2) e_{in}^2 + (1/8) e_{in}^4}{(1-e_{in}^2)^5}  \right].
\end{eqnarray}

Here $\dot{l}_{in} = 2 \pi / P_{in} = ( G (m_1 + m_2) / a_{in}^3 )^{1/2}$ is the mean motion.  Similar expressions with subscripts $1$ and $2$ swapped hold for the second body.  The parameters $(X,Y,Z)$ form a vector in the $( {\bf \hat{e}}_{in},  {\bf \hat{q}}_{in},  {\bf \hat{h}}_{in} )$ frame that gives its angular precession rate relative to the inertial frame.  Their relation to the orbital elements is \citep{1998EKH}:

\begin{eqnarray}
X &=& \dot i \cos \omega_{in} + \dot \Omega_{in} \sin \omega_{in} \sin i \\
Y &=& -\dot i \sin \omega_{in} + \dot \Omega_{in} \cos \omega_{in} \sin i \\
Z &=& \dot \omega_{in} + \dot \Omega_{in} \cos i.
\end{eqnarray}
The dissipationless terms in these parameters can be found by computing $\dot i$, $\dot \omega_{in}$ and $\dot \Omega_{in}$ with the Hamiltonians of \S\ref{analsect}.  For dissipation in $m_1$, the tidal-friction timescale is defined in terms of the viscous timescale $t_{V1}$ (which we take to be a constant):
\begin{equation}
t_{F1} = \frac{t_{V1}}{9} \left( \frac{a_{in}}{R_1} \right)^8 \frac{m_1^2}{(m_1+m_2) m_2} (1+2k_1)^{-2}.
\end{equation}
Here, $k_1$ is the classical apsidal motion constant, a measure of quadrupolar deformability which is related to the Love number ($k_L = 2 k_1$) and the coefficient $Q_E$ given by \cite{2001EK}: $k_{1} = \onehalf Q_E/(1-Q_E)$.  We use the typical value $k_1 = 0.014$ valid for $n=3$ polytropes when representing stars \citep{2001EK} and $k_1 = 0.25$ valid for $n=1$ polytropes when representing gas giant planets.  Again, there is an analogous equation for $k_2$.  This viscosity causes the tidal bulge to lag the instantaneous direction of the companion by a constant time.  In the \cite{1966GS} theory of tides, a quality factor $Q$ is taken to be a constant.  The average fraction of the energy in the tide that is lost to frictional heat per radian of the orbit is $Q^{-1}$, and the tidal bulge lags the instantaneous direction of the companion by an angle $(2 Q)^{-1}$.  When relating that angle to the time lag given by \cite{1998EKH}, taking into account a stray factor of $2$ mentioned in the appendix of \cite{2001EK}, we find:

\begin{equation}
Q = \frac{4}{3} \frac{ k_1 }{ (1+2 k_1)^2} \frac{G m_1}{R_1^3}  \frac{t_{V1}}{\dot{l}_{in}},
\end{equation}
assuming the dissipation is dominant in body $1$.  Taking $t_V$ to be a constant means $Q \propto P_{in}$.


\end{document}